\documentclass[sigconf, authorversion]{acmart}
\AtBeginDocument{%
  }

\copyrightyear{2026}
\acmYear{2026}
\setcopyright{cc}
\setcctype{by}
\acmConference[CHI EA '26]{Extended Abstracts of the 2026 CHI Conference on Human Factors in Computing Systems}{April 13--17, 2026}{Barcelona, Spain}
\acmBooktitle{Extended Abstracts of the 2026 CHI Conference on Human Factors in Computing Systems (CHI EA '26), April 13--17, 2026, Barcelona, Spain}
\acmDOI{10.1145/3772363.3798916}
\acmISBN{979-8-4007-2281-3/2026/04}

\usepackage{subcaption}
\usepackage{array} 
\usepackage{svg}
\usepackage{cleveref} 

\begin{document}

\title{Steering Generative Models for Accessibility: EasyRead Image Generation}

\author{Nicolas Dickenmann}
\authornote{Equal contribution. Correspondence to: ndickenmann@ethz.ch}
\orcid{0009-0007-0370-8417}
\affiliation{%
  \institution{Department of Computer Science \\ ETH Zurich, Switzerland}
  \city{}
  \state{}
  \country{}
}
\author{Yanis Merzouki}
\authornotemark[1]
\orcid{0009-0003-9130-3671}
\affiliation{
  \institution{Department of Computer Science \\ ETH Zurich, Switzerland}
  \city{}
  \state{}
  \country{}
}
\author{Sonia Laguna}
\orcid{0000-0003-3504-2051}
\affiliation{%
  \institution{Department of Computer Science \\ ETH Zurich, Switzerland}
  \city{}
  \state{}
  \country{}
}
\author{Thy Nowak-Tran}
\orcid{0009-0000-8911-4263}
\affiliation{%
  \institution{Unicef Digital Impact Division \\ Valencia, Spain}
  \city{}
  \state{}
  \country{}
}
\author{Emanuele Palumbo}
\orcid{0009-0006-6034-0041}
\affiliation{%
  \institution{ETH AI Center\\ ETH Zurich, Switzerland}
  \city{}
  \state{}
  \country{}
}
\author{Gerda Binder}
\orcid{0009-0004-8600-560X}
\affiliation{%
  \institution{Unicef Digital Impact Division \\ Valencia, Spain}
  \city{}
  \state{}
  \country{}
}
\author{Julia Vogt}
\orcid{0000-0002-6004-7770}
\affiliation{%
  \institution{Department of Computer Science, ETH Zurich, Switzerland}
  \city{}
  \state{}
  \country{}
}

\renewcommand{\shortauthors}{Dickenmann et al.}
\acmArticleType{Review}
\acmCodeLink{https://github.com/borisveytsman/acmart}
\acmDataLink{htps://zenodo.org/link}
\acmContributions{YM, ND: Methodology; Software; Data Curation; Visualization; Investigation; Writing\\
TNT: Conceptualization; Supervision\\
SLC, EP: Supervision; Methodology\\
JV, GB: Supervision; Project Administration}


\begin{abstract}
EasyRead pictograms are simple, visually clear images that represent specific concepts and support comprehension for people with intellectual disabilities, low literacy, or language barriers. The large-scale production of EasyRead content has traditionally been constrained by the cost and expertise required to manually design pictograms. In contrast, automatic generation of such images could significantly reduce production time and cost, enabling broader accessibility across digital and printed materials. However, modern diffusion-based image generation models tend to produce outputs that exhibit excessive visual detail and lack stylistic stability across random seeds, limiting their suitability for clear and consistent pictogram generation. This challenge highlights the need for methods specifically tailored to accessibility-oriented visual content. In this work, we present a unified pipeline for generating EasyRead pictograms by fine-tuning a Stable Diffusion model using LoRA adapters on a curated corpus that combines augmented samples from multiple pictogram datasets. Since EasyRead pictograms lack a unified formal definition, we introduce an EasyRead score to benchmark pictogram quality and consistency. Our results demonstrate that diffusion models can be effectively steered toward producing coherent EasyRead-style images, indicating that generative models can serve as practical tools for scalable and accessible pictogram production.
\end{abstract}

\ccsdesc[500]{Human-centered computing~Accessibility systems and tools}
\ccsdesc[300]{Human-centered computing~Accessibility design and evaluation methods}
\ccsdesc[300]{Human-centered computing~Accessibility technologies}
\ccsdesc[500]{Computing methodologies~Neural networks}
\ccsdesc[500]{Computing methodologies~Learning latent representations}

\keywords{Diffusion Models, Stable Diffusion, LoRA Finetuning, Cognitive Accessibility, Pictogram Generation, EasyRead Metrics, Visual Communication}





\newcommand\blfootnote[1]{%
  \begingroup
  \renewcommand\thefootnote{}\footnote{#1}%
  \addtocounter{footnote}{-1}%
  \endgroup
}

\maketitle


\section{Introduction}
Accessible communication is a prerequisite for meaningful participation in society. EasyRead \citep{easyread} is a widely adopted accessibility framework designed to support people with intellectual disabilities, cognitive impairments, and limited literacy by presenting information in a simplified and visually supported form. In EasyRead documents, information is conveyed through short, concrete sentences paired with clear pictograms or visual symbols that reinforce meaning and reduce cognitive load. 
Despite their importance, EasyRead-compliant images remain difficult to produce at scale \cite{li2024genaibench,hu2023tifa,huang2023t2i}. Creating sentence-aligned pictograms that are semantically accurate, visually simple, and cognitively accessible typically requires substantial manual effort, domain expertise, and iterative design \citep{smith2023easyread}. As a result, EasyRead content creation remains slow, expensive, and largely inaccessible to many communities, practitioners, and organizations that could benefit from it. While recent advances in generative image models have dramatically lowered the barrier for visual content creation, these models are not designed with cognitive accessibility in mind \citep{diffusionmodelsurvey}. Off-the-shelf text-to-image systems tend to generate visually rich, detailed, and stylistically complex images that conflict with the core principles of EasyRead, such as minimalism, strong contrast, and unambiguous depiction of key concepts \cite{rombach2022ldm,saharia2022imagen,zhang2023controlnet}.

This gap motivates a central question: \textbf{can targeted interventions such as lightweight adaptation on curated datasets overcome the challenges of modern, state-of-the-art models to produce compliant EasyRead pictograms?} 

In this work, we test diffusion-based text-to-image models on the task of EasyRead content generation, and validate the effectiveness of lightweight adaptation
for accessibility-driven pictogram generation. In particular, we show that LoRA finetuning on a curated dataset can boost the model's ability to produce high-quality EasyRead-aligned visuals. This approach preserves the scalability and flexibility of modern diffusion models, while enforcing stylistic and semantic constraints necessary for cognitive accessibility.
A key challenge in this domain is evaluation. Existing EasyRead guidelines are largely qualitative and are typically validated through expert review or user studies, which are costly and difficult to scale. To address this, one key contributon in this work is introducing and formalizing a set of pixel-based and semantic metrics that quantify essential properties of EasyRead imagery, such as visual simplicity, contrast, and semantic clarity. These metrics provide, to our knowledge, the first systematic and reproducible framework for measuring “EasyReadness” in generated images, enabling large-scale evaluation and comparison of generative approaches.

In summary, our key contributions in this work are: \textit{(i)} we formalize EasyRead content generation as an accessibility-oriented text-to-image task and benchmark state-of-the-art diffusion models under its constraints; \textit{(ii)} we introduce the \textit{EasyRead score} (ERS) which encompasses quantitative metrics that reflect the quality of EasyRead images for automatic evaluations; \textit{(iii)} we demonstrate that LoRA finetuning on a curated dataset effectively steers diffusion models toward meeting accessibility constraints, releasing the first open-source pipeline tailored for EasyRead generation and evaluation.

\section{Related Work}

Diffusion models can be efficiently adapted to new visual domains through lightweight finetuning techniques such as LoRA \citep{hu2022lora}, DreamBooth \citep{ruiz2023dreambooth}, and Textual Inversion \citep{gal2022textual}. These methods provide strong style specialization while keeping the base model frozen, though there's been limited application to accessibility-oriented visual design.
Prior work on pictogram and icon generation includes Piconet \citep{zhang2021piconet}, symbolic style simplification \citep{liao2022symbolic}, and more recent generative systems such as ICONATE \citep{zhao2020iconate}, IconGAN \citep{chen2022icon}, and Auto-Icon \citep{feng2021autoicon}. These systems focus on improving aesthetics, semantic consistency, or developer workflows, but do not incorporate cognitive accessibility constraints.
Human–computer interaction research provides insights into how icon style influences interpretability. Studies show that stylistic variations affect recognition accuracy and cognitive load \citep{jin2020iconstyle}, and that aesthetics, complexity, and concreteness jointly shape icon usability \citep{collaud2022designicons}. Work in cognitive accessibility \citep{macleod2021accessible} further emphasizes the need for measurable criteria that promote clarity and ease of understanding.
To the best of our knowledge, no prior work integrates state-of-the-art text-to-image diffusion models with explicit, quantitative metrics tailored to cognitive accessibility, nor provides an open-source, scalable pipeline for EasyRead-aligned pictogram generation.

\section{The EasyRead Framework}

To define a consistent standard for EasyRead pictograms, we establish a framework that goes beyond qualitative accessibility guidelines. This framework consists of two primary components: a curated collection of high-quality pictogram data and a set of evaluation metrics designed to measure simplicity and readability.

\subsection{Datasets}
\label{sec:dataset}

In our proposed pipeline for image generation we use a dataset we curate by combining samples from three publicly available pictogram and icon datasets. 
In particular we combine \textit{(i)} \textbf{OpenMoji}~\citep{openmoji}, an open-source emoji set containing $4{,}295$ vector-based icons with consistent styling but limited semantic context; \textit{(ii)} \textbf{ARASAAC}~\citep{arasaac}, a curated set of pictograms used in augmentative and alternative communication with $11{,}972$ samples; and \textit{(iii)} \textbf{LDS}~\citep{lds}, a set of $927$ pictograms used in accessibility contexts. These datasets differ in style, semantic granularity, and metadata quality. 




\subsection{EasyRead Metrics}
\label{sec:easyread-metrics}
The EasyRead metric suite is designed to quantify the extent to which an image conforms to conventions of ``easy-to-understand pictogram''. Rather than emphasizing semantic correctness alone, the suite focuses on metrics that formalize accessibility-oriented design principles. We define six normalized sub-metrics, each capturing a distinct aspect of EasyRead-style imagery, with higher values indicating stronger conformity. The sub-metrics and their associated design principles are summarized in~\Cref{tab:easyread_combined}.
We define the final \textit{EasyRead Score} (ERS) as follows, 
\begin{equation}
\begin{split}
\text{ERS} &= 0.25\,s_{\text{palette}} + 0.20\,s_{\text{edges}} \\
&\quad + 0.15\,s_{\text{saliency}} + 0.15\,s_{\text{contrast}} \\
&\quad + 0.15\,s_{\text{stroke}} + 0.10\,s_{\text{centering}},
\end{split}
\end{equation}
where each $s \in [0,1]$ is a normalized raw metric. 
A detailed description of the raw metric computations of $s_{\text{palette}}$, $ s_{\text{edges}}$, $s_{\text{saliency}}$, $s_{\text{contrast}},s_{\text{stroke}}$, $s_{\text{centering}}$ and their corresponding
normalization functions is provided in Appendix~\ref{appendix:metrics}, as well as an analysis of the distribution of each score, compared to a baseline image dataset.

\begin{table*}[t] 
\centering
\small
\caption{EasyRead design principles and corresponding metrics.}
\vspace*{-0.25cm}
\label{tab:easyread_combined}
\renewcommand{\arraystretch}{1.15}

\begin{tabular}{l p{0.25\textwidth} p{0.28\textwidth} p{0.22\textwidth}}
\toprule
\textbf{EasyRead design principle} & \textbf{Property description} & \textbf{Metric (score)} & \textbf{Metric description} \\ \midrule
Low visual clutter &
Few colors and minimal visual noise &
Palette Complexity ($s_{\text{palette}}$) &
 Color simplicity \\[0.6ex]

Simple geometry &
Clear shapes with smooth outlines &
Edge Density ($s_{\text{edges}}$) &
 Fine-grained line detail \\[0.6ex]

Clear focus &
One visually dominant object &
Saliency Concentration ($s_{\text{saliency}}$) &
 Concentration of visual attention \\[0.6ex]

Strong separation &
Foreground clearly distinct from background &
Foreground--Background Contrast ($s_{\text{contrast}}$) &
 Perceptual foreground--background separation \\[0.6ex]

Stable layout &
Main content centered and well sized &
Centering Error ($s_{\text{centering}}$) &
 Alignment with the image center \\[0.6ex]

Consistent strokes &
Uniform and readable outline thickness &
Relative Stroke Thickness ($s_{\text{stroke}}$) &
 Outline consistency and readability \\

\bottomrule
\end{tabular}
\end{table*}


\section{Method}
In the following section, we present our proposed EasyRead image generation pipeline \footnote{
The code is available at \url{https://github.com/easyread-dsl/easyread_project.git}.
}, articulated in two steps. 
\paragraph{Captioning and Augmentation} The textual metadata in all three datasets lacked sufficient detail for diffusion training, as titles and keywords fail to capture visual attributes such as object count, poses, colors, or inter-object relationships. To address this, we generate natural-language descriptions for each image using the BLIP (Large) Image Captioning model \citep{blip}, producing prompts directly from raw pixels independently of the original labels. Additionally, we augment the ARASAAC dataset \citep{arasaac} by exploiting the unique
customization features  provided through an API on their website. Each pictogram can be rendered with different background colors, skin tones, and hair colors\footnote{The terminology used in our pipeline is inherited from ARASAAC and includes legacy labels that may be considered outdated by current standards. See the Appendix \ref{app:further_details} for further details.}. We systematically vary these parameters during training, enabling the user to control these properties at inference time.
\begin{figure}[ht!]
    \centering
    \includegraphics[width=\columnwidth]{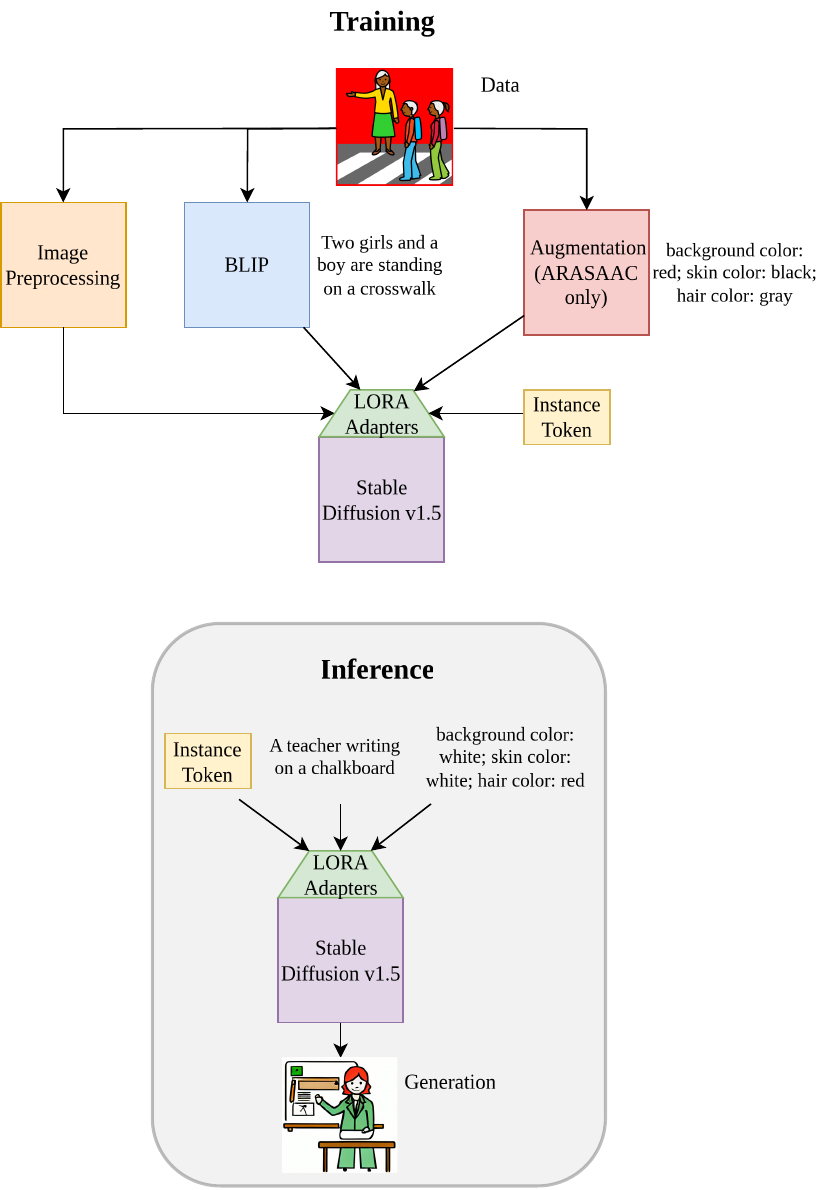}
    \caption{Our pipeline to generate EasyRead pictograms. \textbf{Training (Left)}: Input data is preprocessed, captioned using BLIP and augmented before fine-tuning Stable Diffusion v1.5 using Low-Rank Adaptation (LoRA). A unique instance token (sks) is used to learn the EasyRead style. \textbf{Inference (Right)}: The model generates new pictograms by combining the learned style token (sks) with a descriptive prompt and specific color constraints to ensure stylistic consistency.}
    \label{fig:main}
    \Description{Flowchart of the EasyRead pictogram generation pipeline divided into Training and Inference phases. The Training section illustrates sample data diverging into three parallel steps: Image Preprocessing, BLIP (outputting the caption "Two girls and a boy are standing on a crosswalk"), and ARASAAC-only augmentation (outputting "background color: red; skin color: black; hair color: gray"). These three outputs and an Instance Token are using in training LORA Adapters stacked atop Stable Diffusion version 1.5. The Inference section demonstrates an Instance Token, a text prompt ("A teacher writing on a chalkboard"), and color constraints ("background color: white; skin color: white; hair color: red") feeding directly into the identical model stack to output a matching generated pictogram.}
\end{figure}
\paragraph{Model} Once the data is curated, we fine-tune Stable Diffusion v1.5 to generate EasyRead-style pictograms. The model follows the standard latent diffusion framework, comprising a UNet-based denoising network \citep{unet} conditioned on text embeddings produced by a pretrained CLIP \citep{clip} text encoder, and a variational autoencoder (VAE) \citep{vae} that encodes images into a latent space and decodes latents back to pixel space. During both training and inference, textual prompts are tokenized and embedded using the frozen CLIP tokenizer and text encoder, which provide the conditioning signal for the diffusion process. To adapt the model to the EasyRead pictogram style while minimizing the number of trainable parameters, we employ Low-Rank Adaptation (LoRA) on the UNet attention layers. Specifically, low-rank update matrices are injected into the query, key, value, and output projection matrices of the cross- and self-attention modules. All original Stable Diffusion weights, including the UNet backbone, VAE, and CLIP components, are kept frozen. This parameter-efficient adaptation enables the model to capture the target pictogram style without full model fine-tuning.

\section{Experiments and Results}
\label{sec:experiments}

\paragraph{Experimental Details and Baselines}
We fine-tune rank-16 adapters with a scaling factor of 16, balancing adaptation capacity and parameter efficiency. Captions are prefixed with a dedicated instance token to enable style activation (Appendix~\ref{appendix:data}). Training is performed for 50 epochs on an NVIDIA T4 GPU with batch size 16 using AdamW \((1 \times 10^{-4})\) and mixed precision. For evaluation, Stable Diffusion v1.5 serves as the primary quantitative baseline. We additionally compare against two concurrent closed-source systems, the Global Symbols model~\citep{globalsymbols} and Nano Banana Pro~\citep{nanobanana}, Google’s state-of-the-art general-purpose image generator. As both are accessible only via APIs or web interfaces, they are considered exclusively in qualitative comparisons.

\paragraph{Evaluation Dataset} To assess the impact of EasyRead LoRA fine-tuning on Stable Diffusion v1.5, we construct a controlled prompt set designed to evaluate both semantic faithfulness and stylistic alignment. The set comprises 55 diverse prompts spanning objects, human activities, abstract concepts, and multi-object scenes (see Appendix~\ref{app:val_dataset}). Each prompt follows the ARASAAC augmentation format and specifies background, skin, and hair attributes, even for prompts without people. For each prompt, we generate images with five random seeds in the base and proposed models, resulting in 275 images per model. The only difference between evaluations is that the fine-tuned model prepends the instance token, whereas Stable Diffusion v1.5 prepends "A pictogram of" to preserve the semantic content of the token.

\paragraph{Evaluation Metrics} We evaluate the fine-tuned model using two complementary metrics. First, semantic accuracy is assessed via CLIP similarity, which quantifies how closely the generated image aligns with the prompt. Second, readability and style fidelity are measured using the ERS. For each metric, we report the average score across all 275 generated images for evaluated models.

\begin{table}[htbp]
    \centering
    \caption{Quantitative comparisons between our model and the Stable Diffusion v1.5 baseline on the EasyRead score and CLIP similarity score are summarized as mean $\pm$ standard deviation over 275 samples.}
    \label{tab:model_scores}
    \begin{tabular}{lcc}
        \toprule
        \textbf{Model} & \textbf{ERS} $\uparrow$ & \textbf{CLIP} $\uparrow$ \\
        \midrule
        SD v1.5 (GT) & $0.40 \pm 0.07$ & $24.33 \pm 2.75$ \\
        \textbf{Ours (LoRA)} & $\mathbf{0.47 \pm 0.06}$ & $\mathbf{31.15 \pm 2.88}$ \\
        \bottomrule
    \end{tabular}
\end{table}

\paragraph{Results}~\Cref{tab:model_scores} summarizes the quantitative results. The fine-tuned model outperforms in both EasyRead and CLIP similarity scores: the ERS increase indicates that generated images are stylistically simpler, and better aligned with the design principles; while the improvement in CLIP similarity demonstrates better semantic alignment with the given prompt. These results indicate that a small, targeted LoRA adaptation is sufficient to substantially enhance both stylistic fidelity and semantic accuracy for this task.

\section{Discussion}

\begin{table*}[t] 
    \centering
    \caption{Qualitative single-shot comparison of baseline Stable Diffusion v1.5, our model and external closed-source models (Global Symbols [7] and Nano Banana Pro [8]) on identical prompts.}
    \label{fig:combined_full}
    
    \begin{tabular}{
        >{\centering\arraybackslash}m{0.1\linewidth}
        >{\centering\arraybackslash}m{0.1\linewidth}
        >{\centering\arraybackslash}m{0.1\linewidth}
        >{\centering\arraybackslash}m{0.18\linewidth}
        >{\raggedright\arraybackslash}m{0.42\linewidth}
    }
    \toprule
    \textbf{GT (SD v1.5)} &
    \textbf{Ours (LoRA)} &
    \textbf{Global Symbols} &
    \textbf{Nano Banana Pro} &
    \textbf{Prompt (without sks token)} \\
    \midrule

    \includegraphics[width=\linewidth]{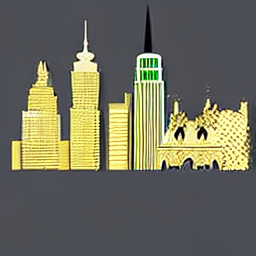} &
    \includegraphics[width=\linewidth]{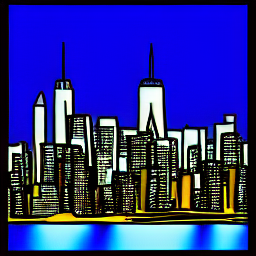} &
    \includegraphics[width=\linewidth]{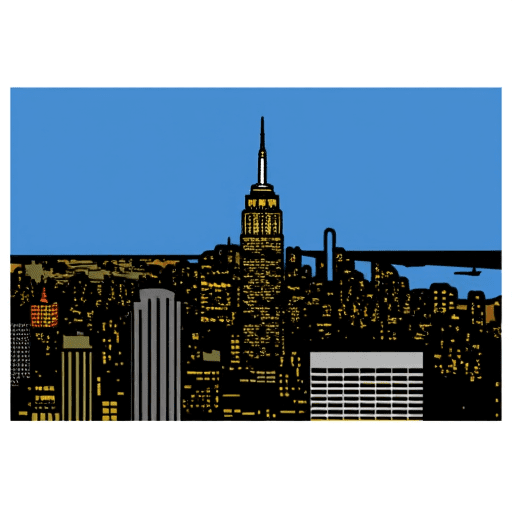} &
    \includegraphics[width=\linewidth]{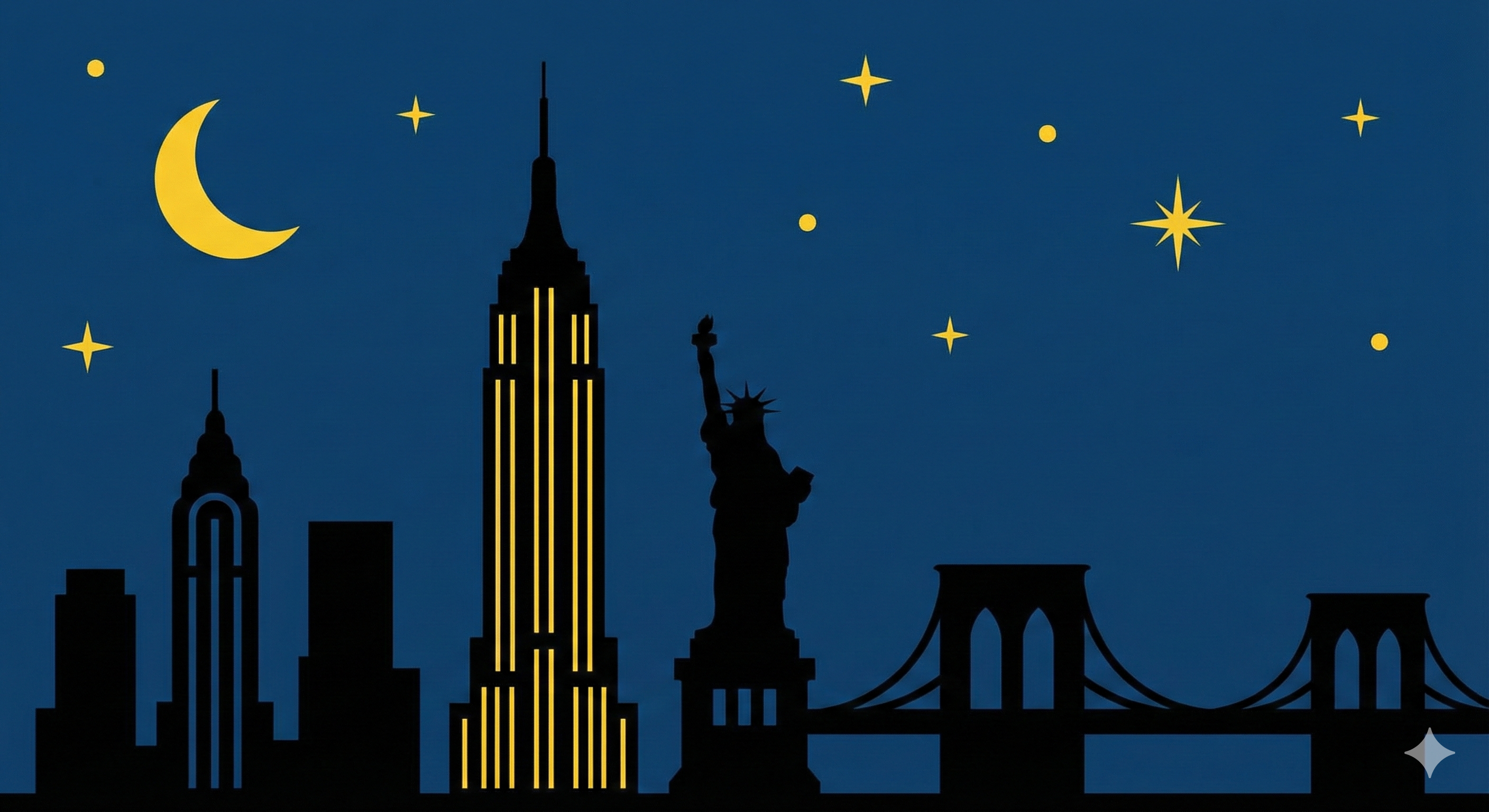} &
    \texttt{The skyline of New York City at night; background color: black; skin color: black; hair color: blonde} \\
    \addlinespace 

    \includegraphics[width=\linewidth]{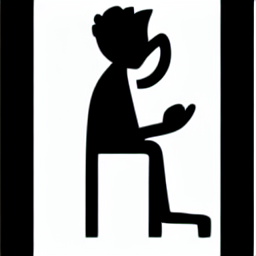} &
    \includegraphics[width=\linewidth]{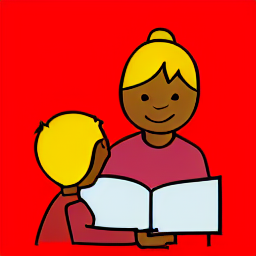} &
    \includegraphics[width=\linewidth]{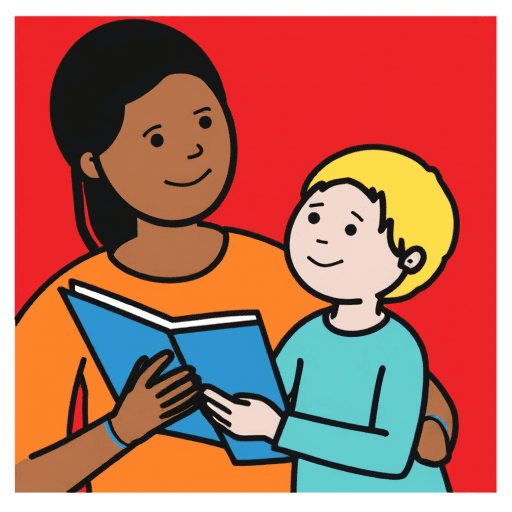} &
    \includegraphics[width=\linewidth]{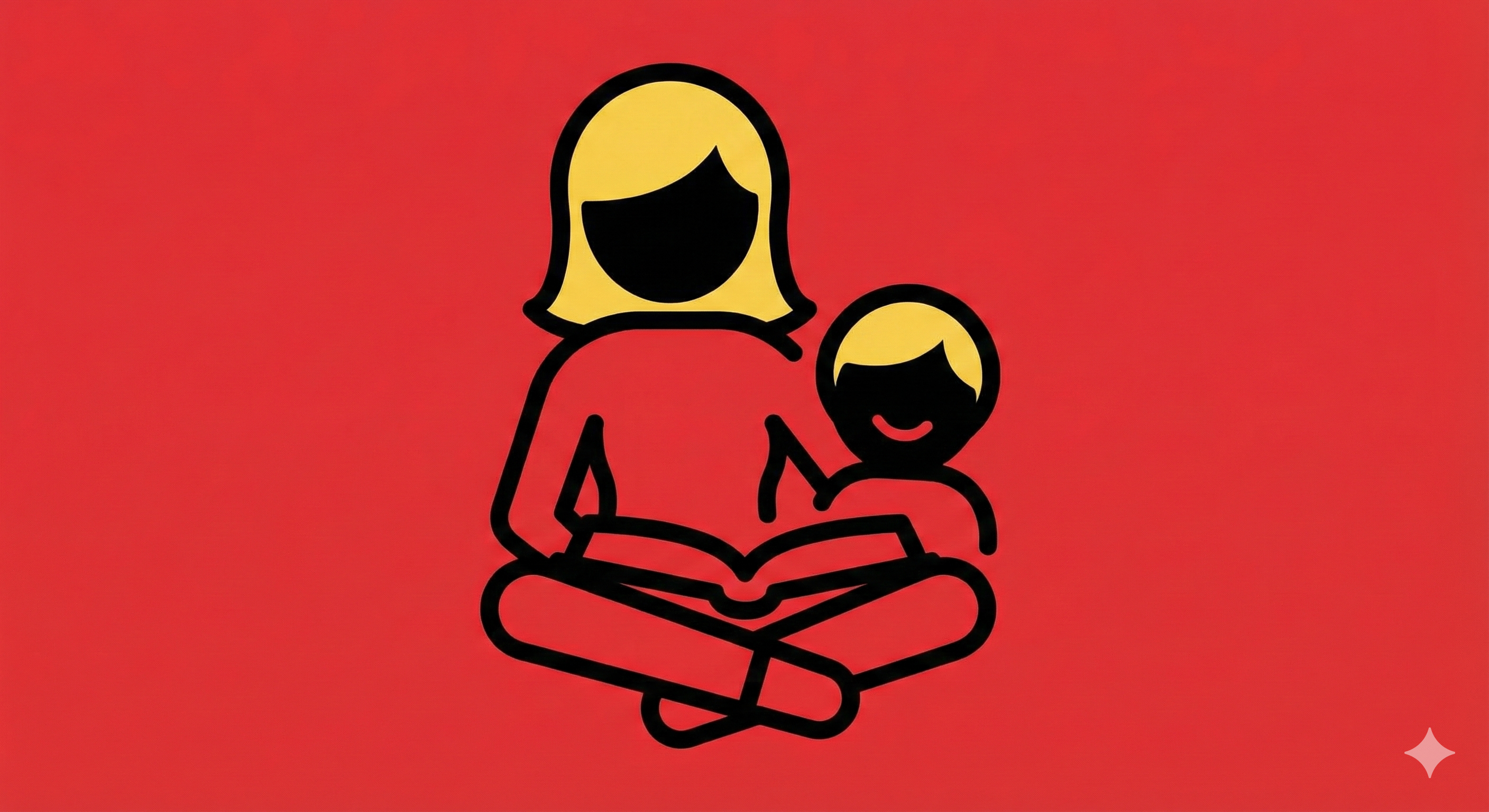} &
    \texttt{A mom reading to a happy child; background color: red; skin color: black; hair color: blonde} \\
    \addlinespace

    \includegraphics[width=\linewidth]{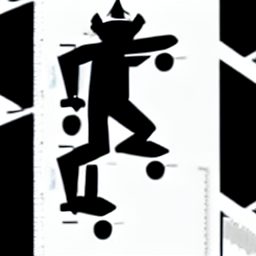} &
    \includegraphics[width=\linewidth]{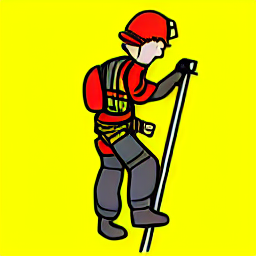} &
    \includegraphics[width=\linewidth]{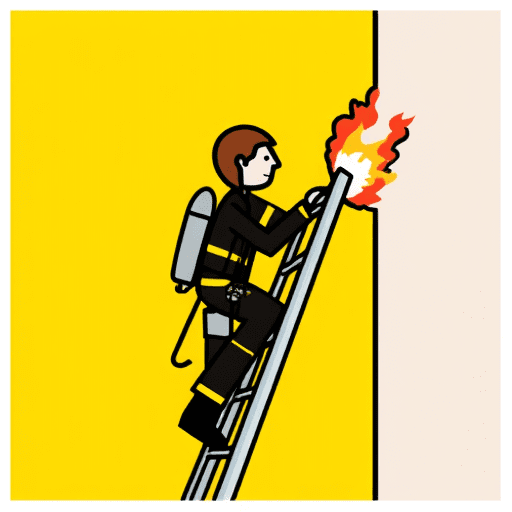} &
    \includegraphics[width=\linewidth]{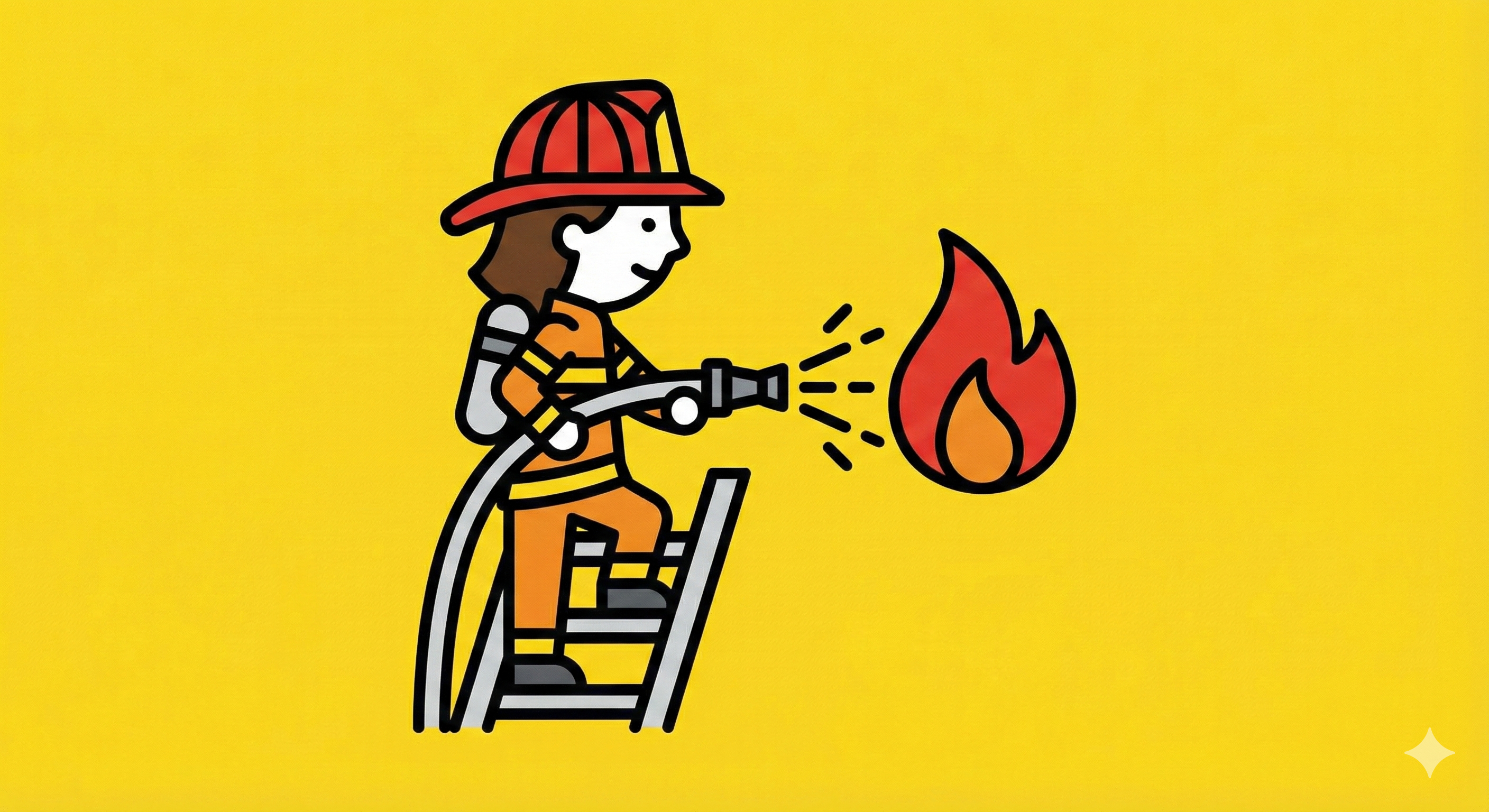} &
    \texttt{A firefighter on a ladder extinguishing a fire; background color: yellow; skin color: white; hair color: brown} \\
    
    \bottomrule
    \end{tabular}
\end{table*}

\begin{figure}[ht] 
    \centering
    \includegraphics[width=0.48\linewidth]{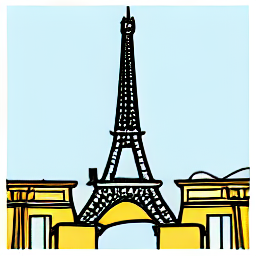}\hfill
    \includegraphics[width=0.48\linewidth]{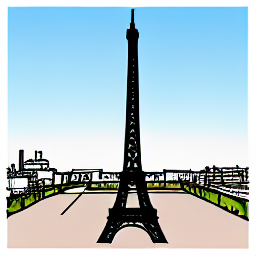}
    
    \vspace{0.3em} 
    
    \includegraphics[width=0.48\linewidth]{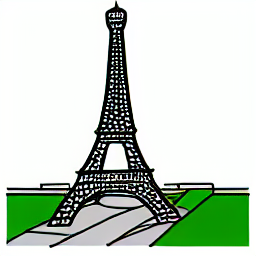}\hfill
    \includegraphics[width=0.48\linewidth]{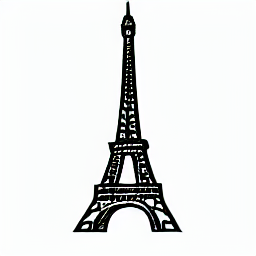}

    \caption{Qualitative sample of four generations of our model at different seeds with the prompt: \texttt{The eifeltower in Paris; background color: white; skin color: white; hair color: black}.}
    \label{fig:2x2grid}
    \Description{Four generated images of the Eiffel Tower showing variations from the same text prompt. The images are arranged in a 2 by 2 grid. Despite the prompt explicitly requesting a white background, the outputs vary significantly:
    \begin{itemize}
        \item \textbf{Top-left:} A stylized, colorful drawing of the tower situated above a yellow archway against a light blue sky.
        \item \textbf{Top-right:} A sketch-style, distant view of the tower at the end of a path with buildings in the background and a blue sky.
        \item \textbf{Bottom-left:} A drawing of the tower standing on a gray path bordered by bright green grass, set against a plain white sky.
        \item \textbf{Bottom-right:} A simple black outline of the tower isolated on a pure white background, being the only image that strictly adheres to the background color constraint.
    \end{itemize}
    }
\end{figure}

\paragraph{Qualitative Evaluation}
Table~\ref{fig:combined_full} presents a qualitative comparison between our fine-tuned model and vanilla Stable Diffusion v1.5. The results demonstrate clear improvements in pictogram generation over baseline Stable Diffusion. Our model consistently follows specified attributes such as background, skin tone, and hair color, and reliably generates human figures and faces in simple scenes.
The table also includes a qualitative comparison with state-of-the-art commercial pictogram generation models. Due to their closed-source nature quantitative evaluation is not feasible, and we therefore restrict the comparison to visual inspection of single-shot generations. Rather than making definitive performance claims, we report observed trends: our model often produces outputs that are comparable to, and in some cases visually more consistent than, these commercial systems, especially in terms of style uniformity and adherence to color and attribute instructions. By contrast, Nano Banana Pro, a strong general-purpose image generator, is not specialized for pictogram generation and occasionally exhibits style inconsistencies or partial instruction compliance (additional examples are provided in Appendix~\ref{app:further_samples}).
We emphasize that these comparisons are qualitative and illustrative, and serve to contextualize our model’s behavior rather than to establish a formal ranking against proprietary systems.

\paragraph{Output Variability and Instruction Following} Figure \ref{fig:2x2grid} demonstrates that the model exhibits noticeable output variability across different random seeds. This variability is handled by generating four samples per prompt, from which a preferred output can be selected. Despite this stochasticity, the model follows instructions reliably: it consistently disregards attributes that are not applicable (e.g., skin or hair color when no human figure is present), indicating that such constraints have been effectively learned during training. Background-related instructions are followed with slightly more flexibility. In some cases, the model augments the scene with coherent contextual elements such as a blue sky (top-row samples), while in others it adheres strictly to the specified background color (bottom-row samples).

\paragraph{Limitations and Future Work} Our model exhibits difficulty preserving details in complex scenes, such as the firefighter example in Table \ref{fig:combined_full}. This limitation is particularly evident when generating multiple human figures, where facial features or other important details are often lost. Additionally, while ARASAAC-based training augmentations allow us to vary skin tone, this does not fully capture ethnicity, as it would involve features beyond skin color. Future research trajectories could integrate differentiable EasyRead components directly into the training objective to explicitly optimize for perceptual clarity. To further enforce stylistic constraints such as simplicity and coherent contours, future lines can also explore the incorporation of style-consistency losses derived from dataset-level Gram statistics. 
Finally, human-centric validation remains a critical frontier and future directions should explore structured user studies to cross-reference the proposed ERS with qualitative human perception. Similarly, the utility of the generation pipeline can be assessed through expert-in-the-loop evaluations of prototype interfaces.

\section{Conclusion}
We present the first open and scalable image generation pipeline explicitly designed for EasyRead and pictogram-style visual communication. By introducing a targeted augmentation strategy for pictogram datasets and integrating BLIP-enhanced captions with LoRA-based fine-tuning of Stable Diffusion, our approach enables controllable generation of accessibility-critical attributes such as background simplicity, skin tone, and hair color while preserving strong generalization to unseen concepts. Crucially, we move beyond purely qualitative evaluation by formalizing automated metrics to quantify “EasyReadness,” enabling systematic measurement and comparison where prior work relied largely on user studies. Our results demonstrate that state-of-the-art diffusion models can be effectively adapted to produce semantically clear, visually simple, and accessible imagery, establishing a practical foundation for inclusive image generation and providing a valuable open framework for accessibility-driven communities and future research.

\begin{acks}
 SL is supported by the Swiss State Secretariat for Education, Research and Innovation (SERI) under contract number MB22.00047. EP is supported by a fellowship from the ETH AI Center, and received funding from the grant \#2021-911 of the Strategic Focal Area “Personalized Health and Related Technologies (PHRT)” of the ETH Domain (Swiss Federal Institutes of Technology).
\end{acks}

\bibliographystyle{ACM-Reference-Format} 
\bibliography{bibliography} 

\clearpage
\appendix

\section{Data}
\label{appendix:data}

\begin{table}[hbt!]
\centering
\setlength{\tabcolsep}{0.2em} 

\renewcommand{\arraystretch}{0.5} 

\begin{tabular}{c | c | c}
    \textbf{ARASAAC} & \textbf{OpenMoji} & \textbf{LDS} \\ 
    \hline
    \rule{0pt}{1.25em} 
    \includegraphics[width=0.31\columnwidth]{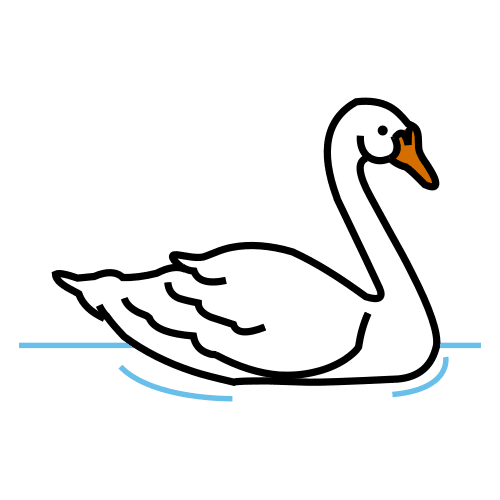} &
    \includegraphics[width=0.31\columnwidth]{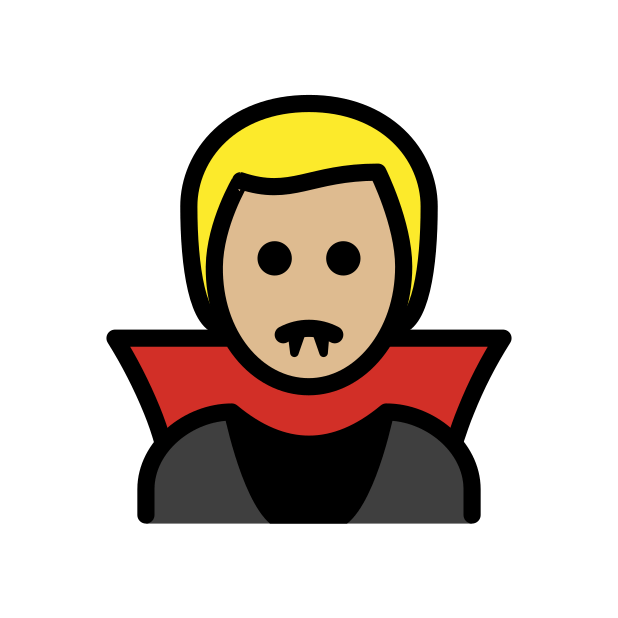} &
    \includegraphics[width=0.31\columnwidth]{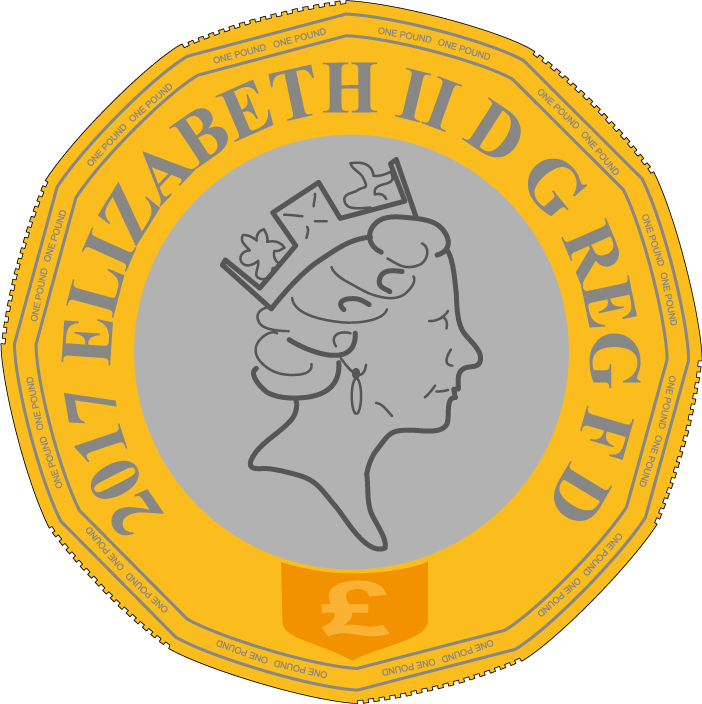} \\[0.2em]
    
    \includegraphics[width=0.31\columnwidth]{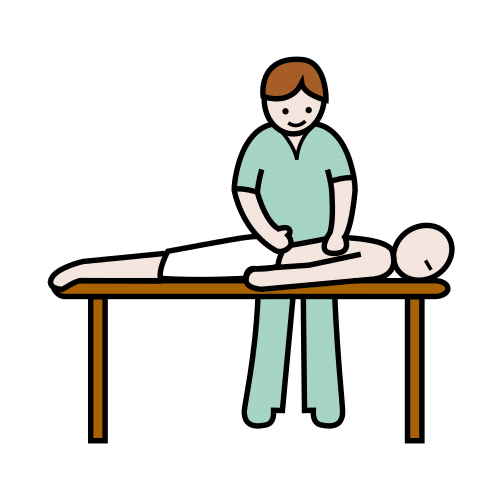} &
    \includegraphics[width=0.31\columnwidth]{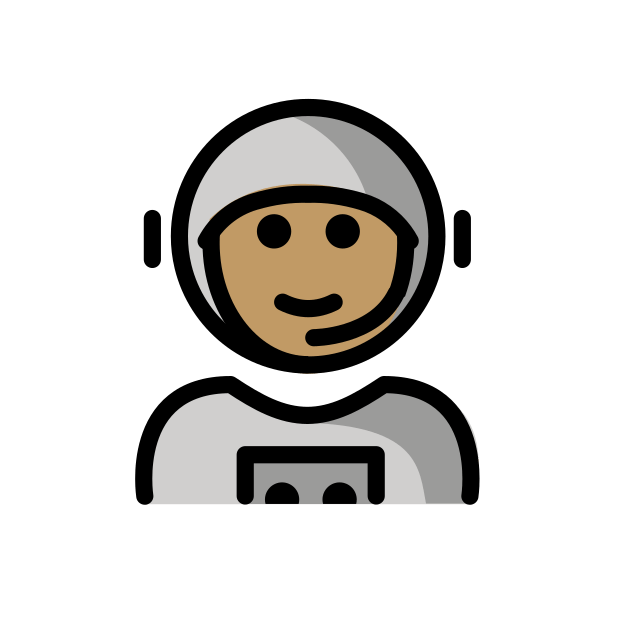} &
    \includegraphics[width=0.31\columnwidth]{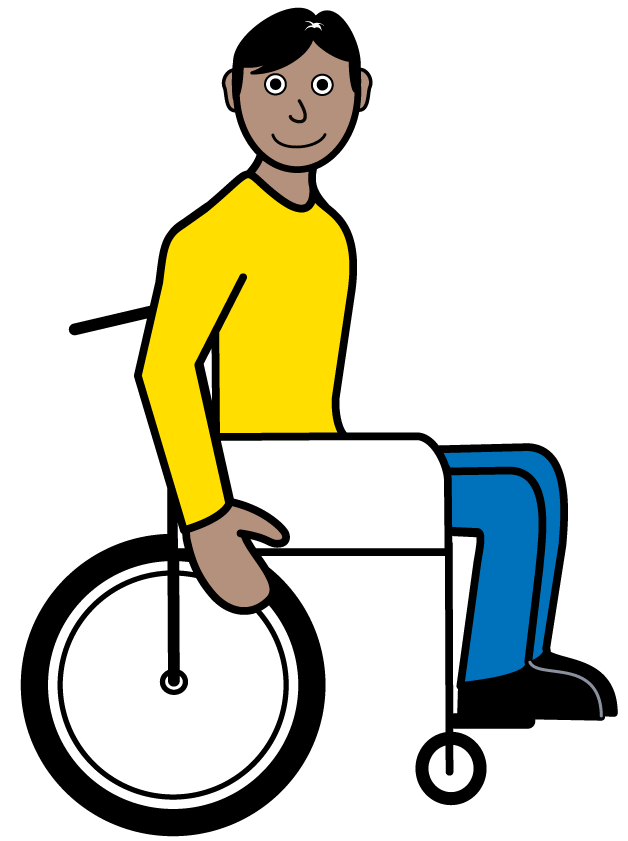} \\[0.2em]
    
    \includegraphics[width=0.31\columnwidth]{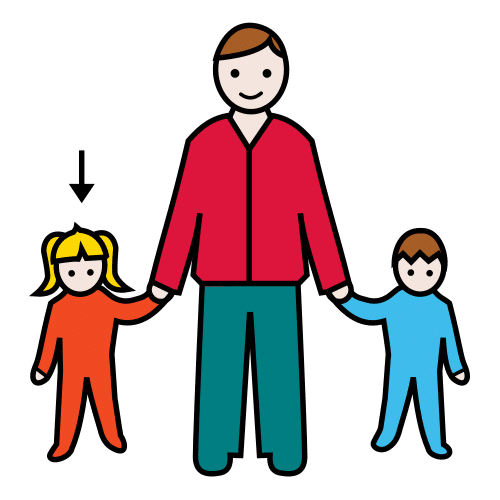} &
    \includegraphics[width=0.31\columnwidth]{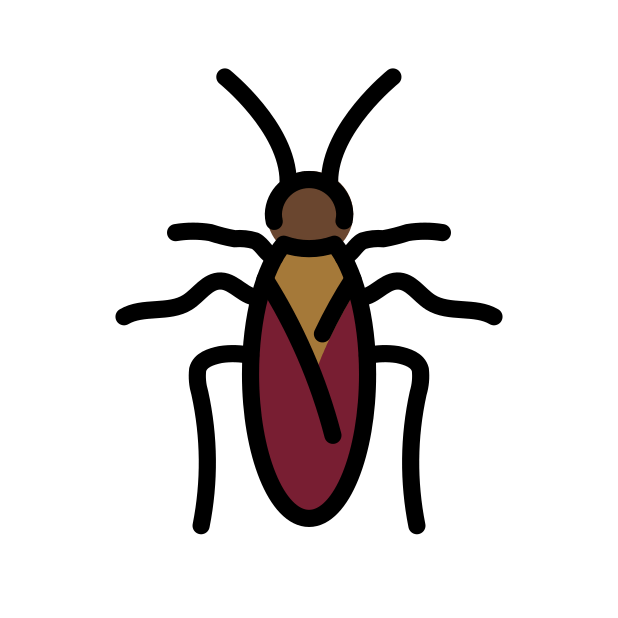} &
    \includegraphics[width=0.31\columnwidth]{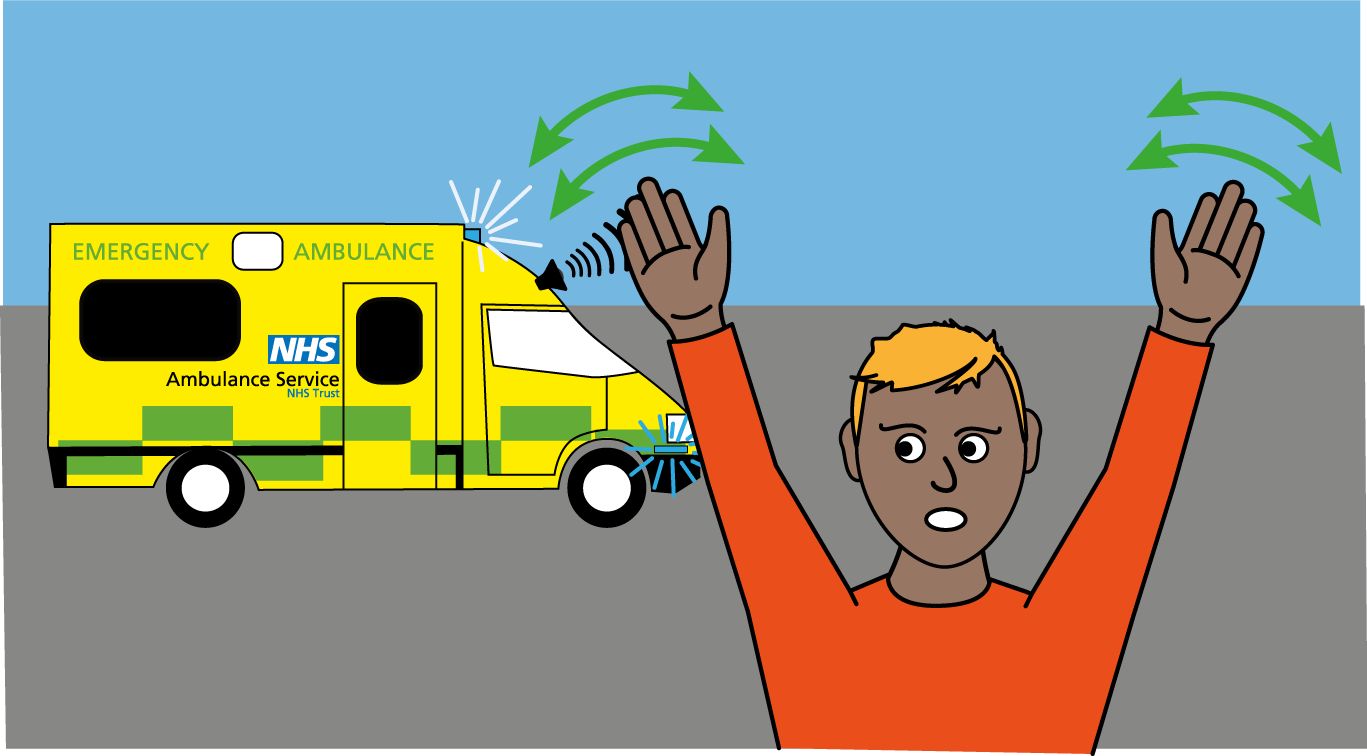} \\
\end{tabular}

\caption{\textbf{Examples of the three datasets in our corpus: ARASAAC pictograms, OpenMoji icons, and LDS symbolic illustrations.}
Although all are simplified drawings, they exhibit clear and systematic stylistic differences in visual abstraction, line weight, and compositional conventions.}
\end{table}

\subsection{Cleaning Generated Captions}
The caption generation pipeline utilizing BLIP
includes several post-processing steps: removing boilerplate prefixes (e.g., ``an
image of''), eliminating the word ``cartoon'' (redundant since all dataset points are icons and cartoon-like images), normalizing whitespace, and
capitalizing the final text. These heuristics encourage clean, EasyRead-compatible
captions that focus on concrete visual content. In \Cref{fig:caption_table_comparison} we present examples of the generated captions.

\subsection{Further Details on the Integration of the ARASAAC Augmentation}
We append each training prompt with "; background color: {background color}; skin color: {skin color};
hair color: {hair color}". In \Cref{fig:arasaaccontrol} we present examples of controllable prompts and their generated samples.
The background color can be varied by HEX code and we chose: red, green, blue, yellow, black, white.
Skin color options were based on the ARASAAC customization tool and included the following categories: White, Black, Asian, Mulatto, and Aztec. These categories are inherited from the ARASAAC customization tool and do not align with current standards for describing skin tone, race, or ethnicity. The labels reflect legacy terminology and function only as internal visual appearance identifiers within the tool’s color-selection interface; they are not user-facing. They are reproduced verbatim solely for consistency with the tool’s predefined options and do not represent the authors’ endorsed terminology. The hair color customizations by ARASAAC are: blonde, brown, darkBrown, gray, darkGray, red, black.
When augmenting the data we rotate through all the properties in order to get all possible combinations. Images without people in it get the labels too as this allows the model to learn ignoring irrelevant properties.
Additionally, we prepend a sks token, a synthetic token, added to encode a novel concept, EasyRead pictograms in our case. During fine-tuning, the model learns to associate the token with EasyRead pictogram rules. At inference time, inserting the SKS token into a text prompt conditions the diffusion process on the learned concept, enabling controlled generation of images that reflect the target concept without retraining the full model. 
\label{app:further_details}
\begin{table}[t]
    \centering
    \begin{minipage}[t]{0.48\textwidth}
        \caption{This table presents the captions generated by BLIP alongside the original labels.}
        \label{fig:caption_table_comparison}
        \centering
        \resizebox{\linewidth}{!}{%
            \begin{tabular}{>{\centering\arraybackslash}m{0.3\linewidth}|>
            {\raggedright\arraybackslash}m{0.2\linewidth}|>{\raggedright\arraybackslash}m{0.35\linewidth}}
                \textbf{Image} & 
                \textbf{Label (original)} & \textbf{BLIP Caption} \\
                \hline
                \rule{0pt}{1.25em}
                \includegraphics[width=1\linewidth]{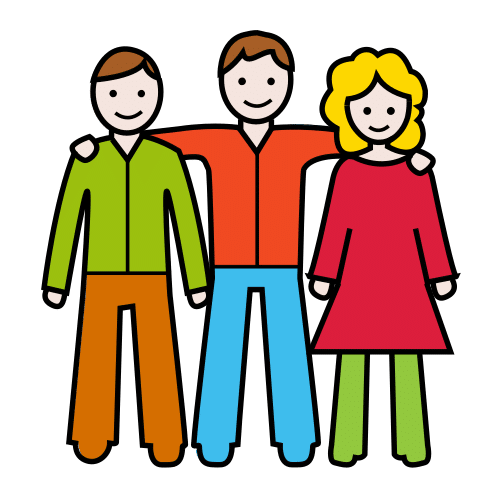} &
                 \texttt{friends} &
                 \texttt{Three people standing together with their arms around each other} \\
                \includegraphics[width=1\linewidth]{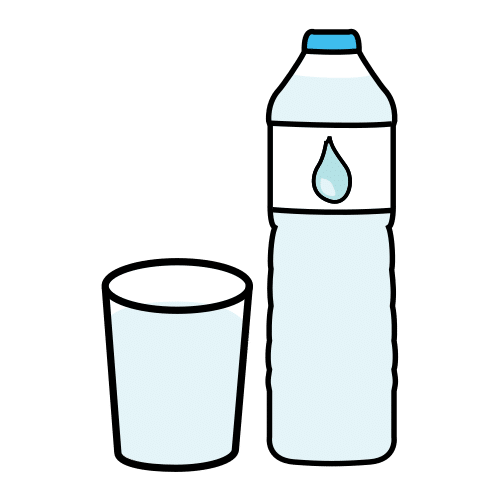} &
                 \texttt{water} &
                 \texttt{A bottle of water next to a glass of water}
            \end{tabular}
        }
    \end{minipage}
    \hfill 
    \begin{minipage}[t]{0.48\textwidth}
        \caption{This table displays the augmentations provided by the ARASAAC dataset that allow for control parameters during training and inference.}
        \label{fig:arasaaccontrol}
        \centering
        \resizebox{\linewidth}{!}{%
            \begin{tabular}{m{0.3\columnwidth}| m{0.3\columnwidth}|m{0.4\columnwidth}} 
                \textbf{\large ARASAAC \mbox{sample}} &
                \textbf{\large augmented ARASAAC \mbox{sample}} & 
                \textbf{\large Training prompt} \\
                \hline
                \rule{0pt}{1.25em}
                \includegraphics[width=0.3\columnwidth]{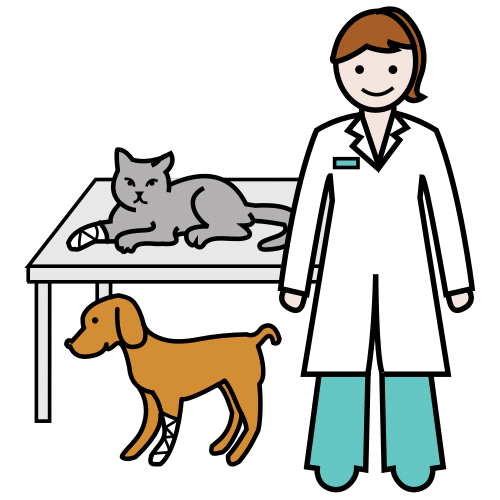} & 
                \includegraphics[width=0.3\columnwidth]{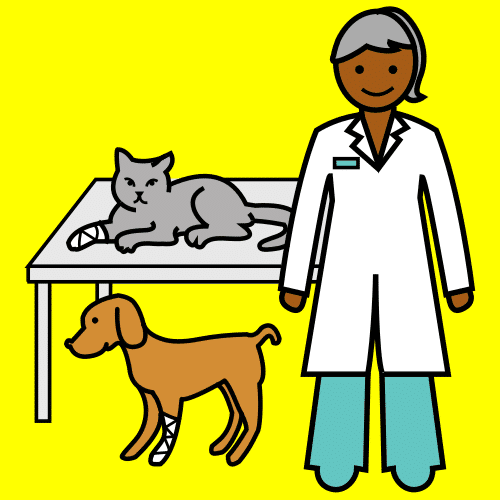} &
                \texttt{veterinary: A vet with a cat and dog on a table; background color: yellow; skin color: black; hair color: darkGray} \\
                \includegraphics[width=0.3\columnwidth]{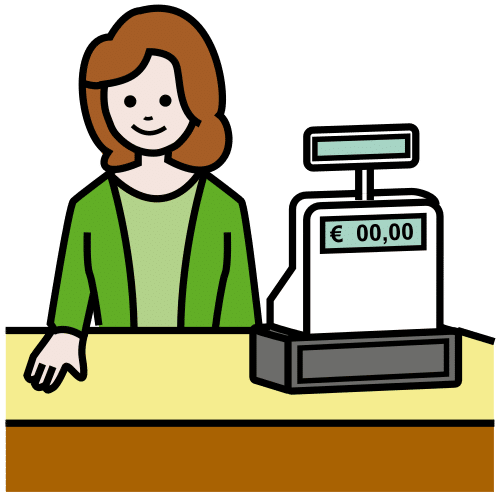} & 
                \includegraphics[width=0.3\columnwidth]{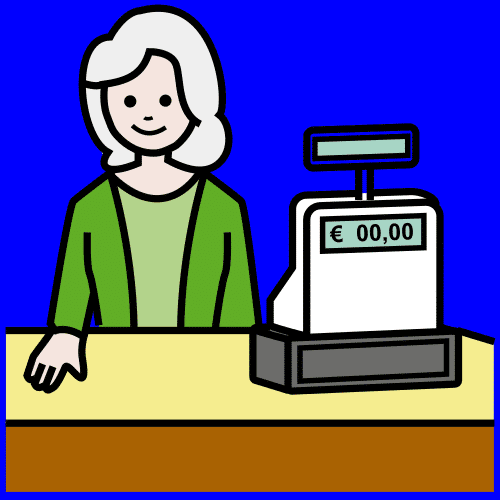} &
                 \texttt{cashier: A woman standing at a counter with a cash register on it; background color: blue; skin color white; hair color: gray} \\
            \end{tabular}
        }
    \end{minipage}
\end{table}

\subsection{Validation Dataset}
\label{app:val_dataset}
We compiled a dataset of 55 prompts to evaluate our model against other baselines using clip similarity and the ERS. Please examine the 55 prompts in Table \ref{tab:prompts}. Moreover, in Appendix \ref{app:further_samples} we present additional qualitative samples of our model generated using the prompts of the validation dataset.

\newpage
{
\begin{table*}[htbp]
\centering
\caption{Validation Dataset prompts and control attributes}
\label{tab:prompts}
\small 
\renewcommand{\arraystretch}{0.97}
\begin{tabular}{p{7cm} l l l}
\toprule
\textbf{Scene Description} & \textbf{Background} & \textbf{Skin} & \textbf{Hair} \\
\midrule

The skyline of NYC at night & Red & White & Blonde \\
A person watering a houseplant & Green & Black & Brown \\
A child brushing their teeth & Blue & Asian & Dark Brown \\
A doctor listening to heartbeat & Yellow & Mulatto & Gray \\
A firefighter climbing a ladder & Black & Aztec & Dark Gray \\
A teacher writing on a chalkboard & White & White & Red \\
A bicycle leaning against a wall & Red & Black & Black \\
A dog catching a frisbee & Green & Asian & Blonde \\
An airplane taking off & Blue & Mulatto & Brown \\
A person cooking soup & Yellow & Aztec & Dark Brown \\
A person reading a map outdoors & Black & White & Gray \\
A mom helping child with homework & White & Black & Dark Gray \\
A swimmer diving into a pool & Red & Asian & Red \\
A person running in the rain & Green & Mulatto & Black \\
A chef preparing vegetables & Blue & Aztec & Blonde \\
A person repairing a bicycle tire & Yellow & White & Brown \\
A commuter waiting for a bus & Black & Black & Dark Brown \\
A dentist examining a patient & White & Asian & Gray \\
A person opening a window & Red & Mulatto & Dark Gray \\
A delivery worker carrying package & Green & Aztec & Red \\
A person folding laundry & Blue & White & Black \\
A person feeding a baby & Yellow & Black & Blonde \\
A person practicing yoga & Black & Asian & Brown \\
A police officer directing traffic & White & Mulatto & Dark Brown \\
A tree with falling leaves & Red & Aztec & Gray \\
A cup of coffee on a table & Green & White & Dark Gray \\
A car driving on a mountain road & Blue & Black & Red \\
A lighthouse overlooking the ocean & Yellow & Asian & Black \\
A bus picking up passengers & Black & Mulatto & Blonde \\
A person taking out the trash & White & Aztec & Brown \\
A person tying their shoes & Red & White & Dark Brown \\
A construction worker laying bricks & Green & Black & Gray \\
A nurse giving an injection & Blue & Asian & Dark Gray \\
A person using a laptop & Yellow & Mulatto & Red \\
A person painting a wall & Black & Aztec & Black \\
A person holding an umbrella & White & White & Blonde \\
A bird flying above a lake & Red & Black & Brown \\
A soccer player kicking a ball & Green & Asian & Dark Brown \\
A person waiting at a crosswalk & Blue & Mulatto & Gray \\
A person fishing from a boat & Yellow & Aztec & Dark Gray \\
A person giving a presentation & Black & White & Red \\
A person carrying groceries & White & Black & Black \\
A cat sleeping on a cushion & Red & Asian & Blonde \\
A train arriving at a station & Green & Mulatto & Brown \\
A person washing their hands & Blue & Aztec & Dark Brown \\
A person climbing stairs & Yellow & White & Gray \\
A person chopping vegetables & Black & Black & Dark Gray \\
A person taking a photograph & White & Asian & Red \\
A street market with food stands & Red & Mulatto & Black \\
A person putting on a jacket & Green & Aztec & Blonde \\
A person calling for help & Blue & White & Brown \\
A person holding a first aid kit & Yellow & Black & Dark Brown \\
A person turning off a light switch & Black & Asian & Gray \\
A person cleaning a window & White & Mulatto & Dark Gray \\
A mom reading to a happy child & Red & Black & Blonde \\
\bottomrule
\end{tabular}
\end{table*}

\clearpage
\section{Further Samples}
\label{app:further_samples}
In \Cref{fig:further_sample} we present further results generated by our model on prompts from the validation dataset (see Appendix \ref{app:val_dataset}). In \Cref{fig:nanobanana} we present four samples with the same prompt at different seeds generated with closed-source Nano Banana Pro \citep{nanobanana}. We observe that fine-tuning Stable Diffusion v1.5 with an sks-instance token yields a persistent style across generations, while Nano Banana Pro samples exhibit substantially greater stylistic variability across seeds, including differences in background, typography, and perspective.

\begin{figure}[htbp]
    \centering
    \begin{minipage}{0.95\columnwidth} 
        \centering
        \includegraphics[width=0.48\linewidth]{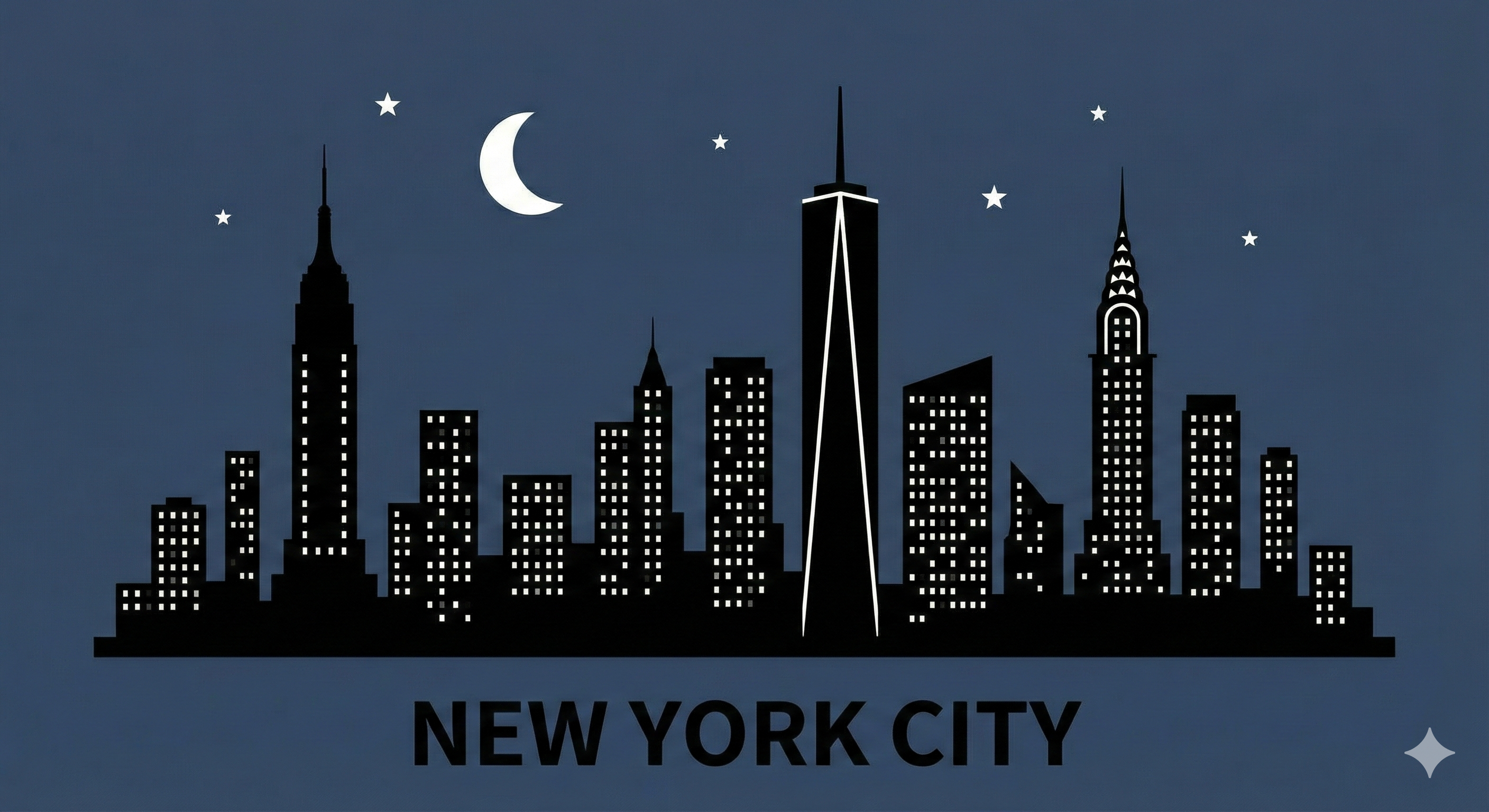}\hfill
        \includegraphics[width=0.48\linewidth]{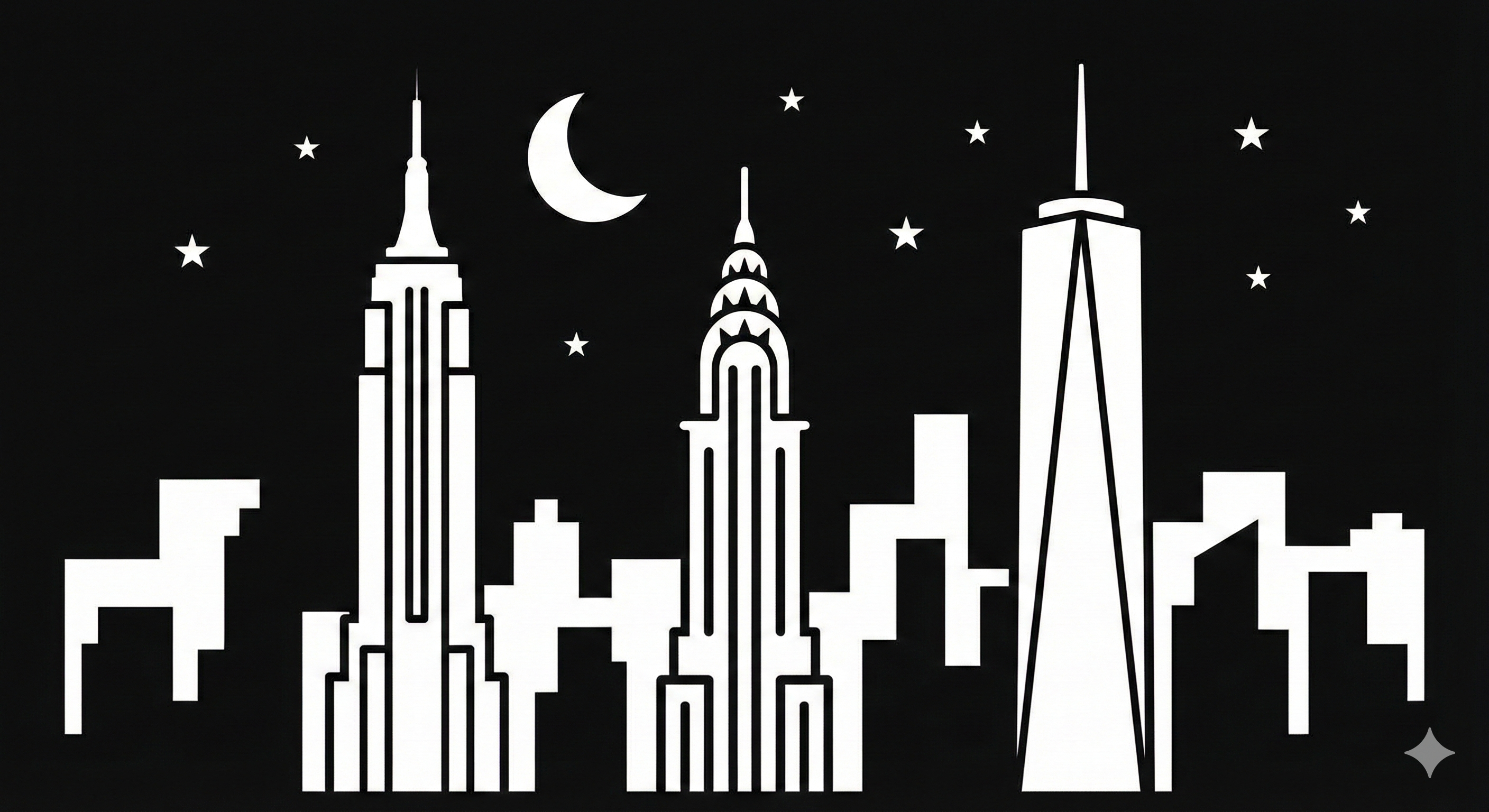}
        
        \vspace{0.3em}
        
        \includegraphics[width=0.48\linewidth]{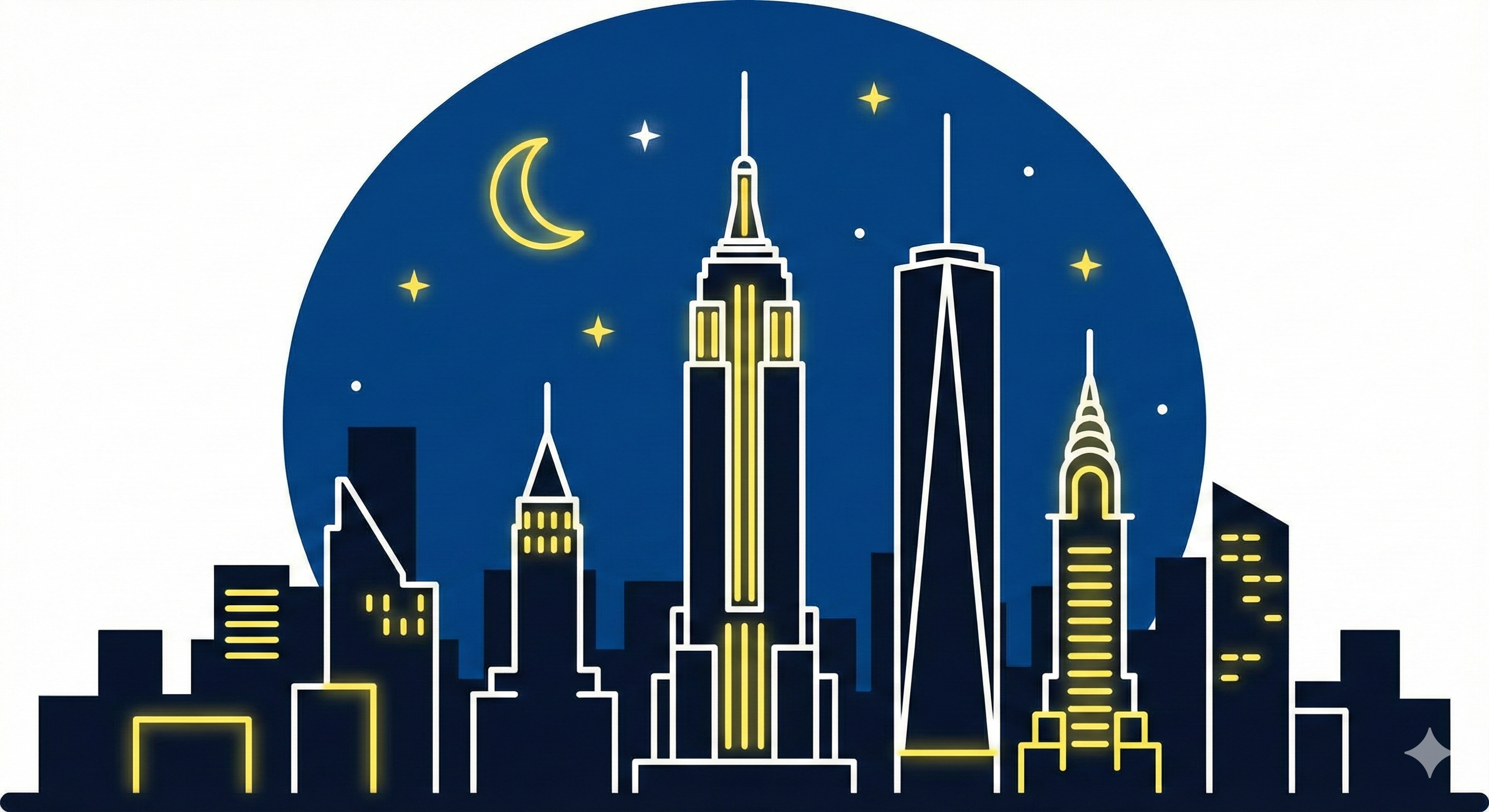}\hfill
        \includegraphics[width=0.48\linewidth]{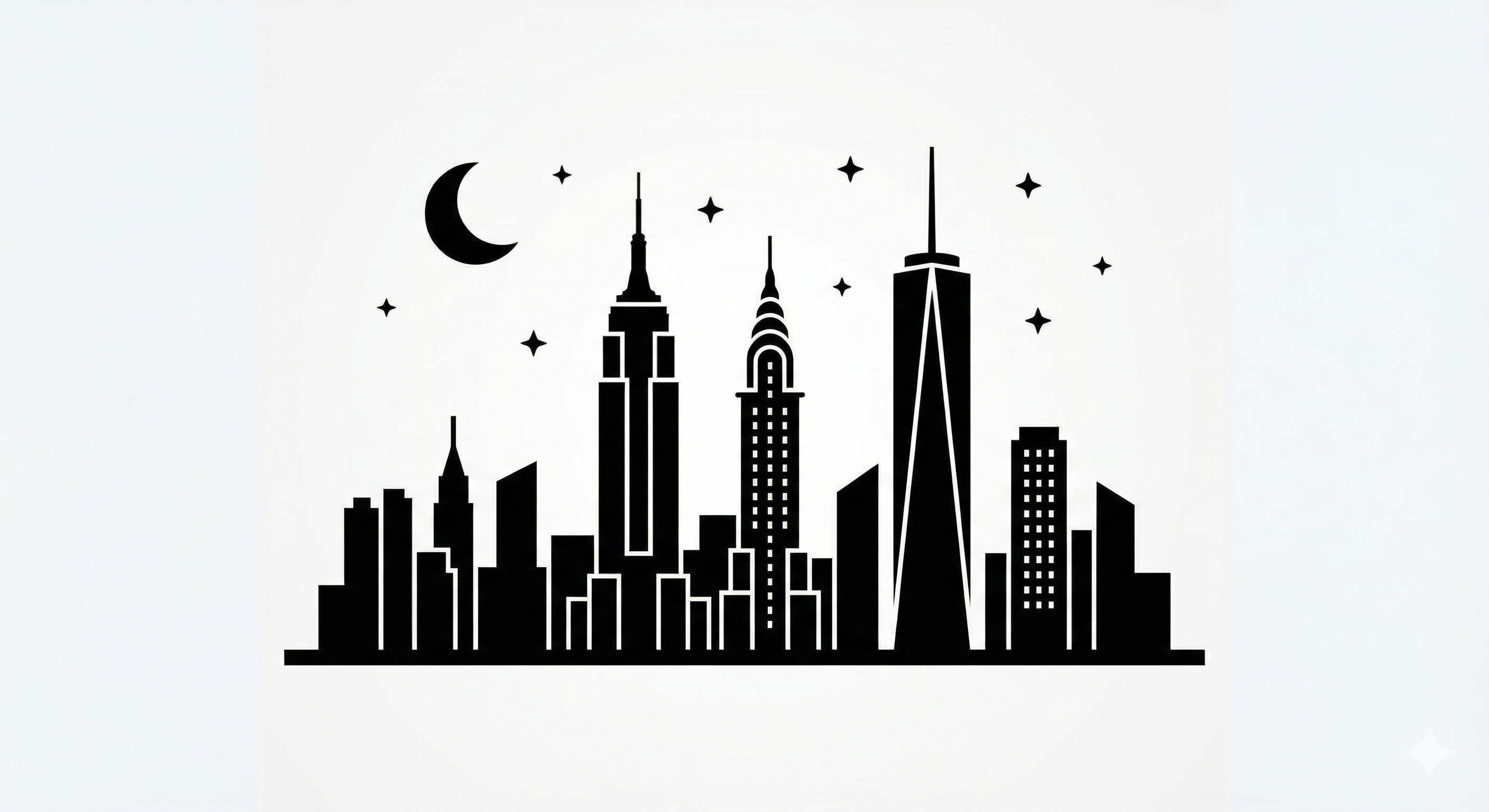}

        \caption{Four samples at different seeds generated by the closed-source Nano Banana Pro \citep{nanobanana} model with the prompt: \texttt{A pictogram of the skyline of New York City at night}.}
        \Description{A 2 by 2 grid displaying four different generated pictograms of the New York City skyline at night:
        \begin{itemize}
            \item \textbf{Top-left:} A black silhouette of the skyline against a dark blue background, featuring a white crescent moon and stars, with the words ``NEW YORK CITY'' written in black text below the buildings.
            \item \textbf{Top-right:} A white outline and solid white shapes forming the skyline against a solid black background, with a white crescent moon and stars.
            \item \textbf{Bottom-left:} A skyline with dark silhouettes and yellow glowing accents, framed within a dark blue circular background containing a yellow crescent moon and stars.
            \item \textbf{Bottom-right:} A black silhouette and outline of the skyline against a very pale, off-white background, featuring a black crescent moon and stars.
        \end{itemize}
        }
        \label{fig:nanobanana}
    \end{minipage}
\end{figure}

\begin{table*}[t]
\centering
\footnotesize
\caption{Further samples generated by our model. Five samples at different seeds for each prompt}
\resizebox{0.9\textwidth}{!}{%
\begin{tabular}{
    >{\centering\arraybackslash}m{0.30\textwidth} |
    >{\centering\arraybackslash}m{0.16\textwidth} |
    >{\centering\arraybackslash}m{0.16\textwidth} |
    >{\centering\arraybackslash}m{0.16\textwidth} |
    >{\raggedright\arraybackslash}m{0.16\textwidth} |
    >{\raggedright\arraybackslash}m{0.16\textwidth}
}
\textbf{\large Prompt (without sks token)} &
\textbf{\large Generation 1} &
\textbf{\large Generation 2} &
\textbf{\large Generation 3} &
\textbf{\large Generation 4} &
\textbf{\large Generation 5} \\
\hline
\rule{0pt}{1.25em}

\texttt{A street market with food stands; background color: red; skin color: mulatto; hair color: black} &
\includegraphics[width=\linewidth]{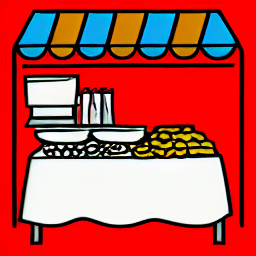} &
\includegraphics[width=\linewidth]{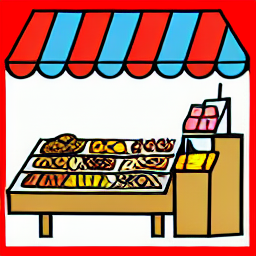} &
\includegraphics[width=\linewidth]{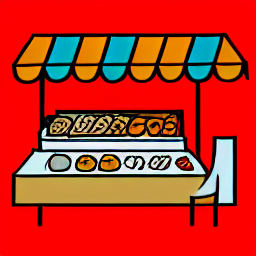} &
\includegraphics[width=\linewidth]{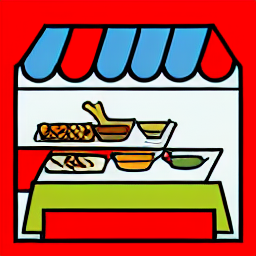} &
\includegraphics[width=\linewidth]{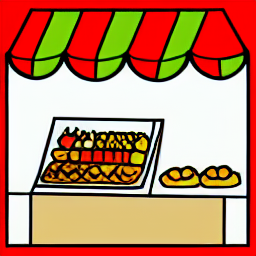}
 \\
 \texttt{A person taking a photograph; background color: white; skin color: asian; hair color: red} &
\includegraphics[width=\linewidth]{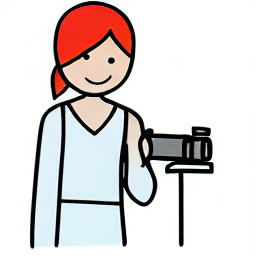} &
\includegraphics[width=\linewidth]{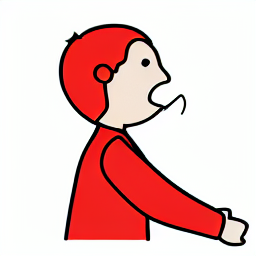} &
\includegraphics[width=\linewidth]{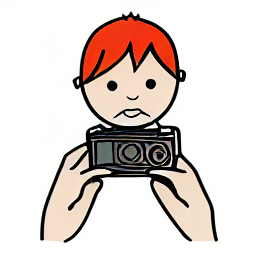} &
\includegraphics[width=\linewidth]{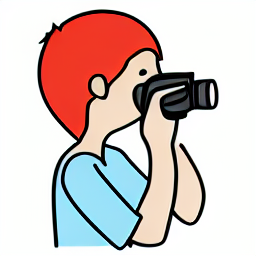} &
\includegraphics[width=\linewidth]{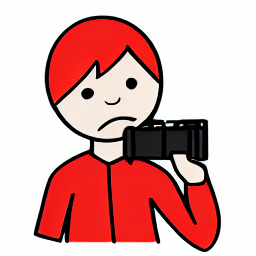}
 \\
  \texttt{A soccer player kicking a ball; background color: green; skin color: asian; hair color: darkBrown} &
\includegraphics[width=\linewidth]{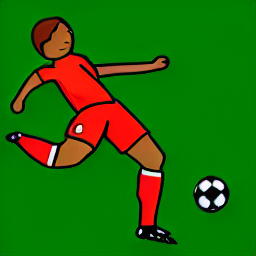} &
\includegraphics[width=\linewidth]{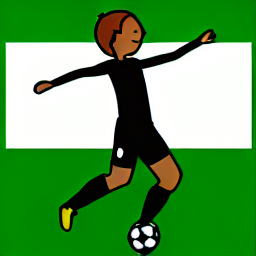} &
\includegraphics[width=\linewidth]{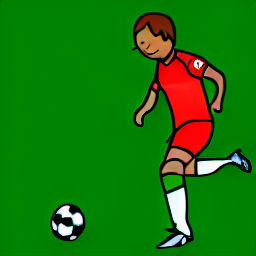} &
\includegraphics[width=\linewidth]{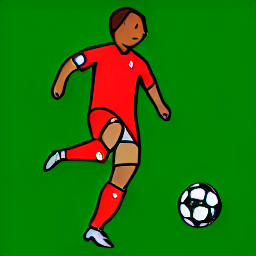} &
\includegraphics[width=\linewidth]{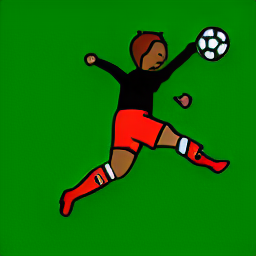}
 \\
   \texttt{A cup of coffee on a table; background color: green; skin color: white; hair color: darkGray} &
\includegraphics[width=\linewidth]{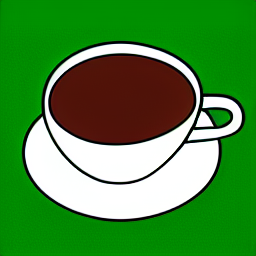} &
\includegraphics[width=\linewidth]{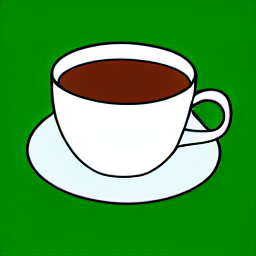} &
\includegraphics[width=\linewidth]{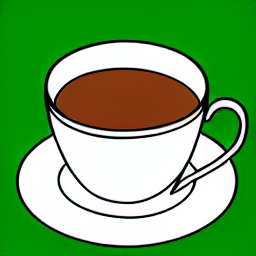} &
\includegraphics[width=\linewidth]{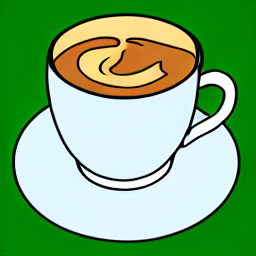} &
\includegraphics[width=\linewidth]{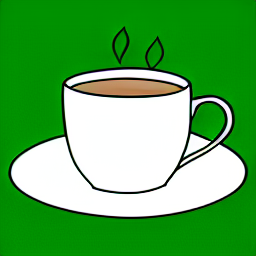}
 \\
   \texttt{A police officer directing traffic; background color: white; skin color: mulatto; hair color: darkBrown} &
\includegraphics[width=\linewidth]{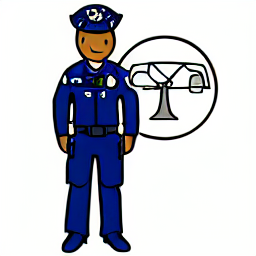} &
\includegraphics[width=\linewidth]{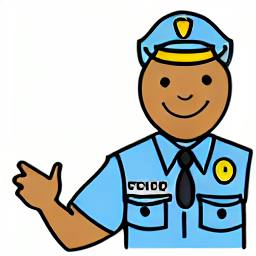} &
\includegraphics[width=\linewidth]{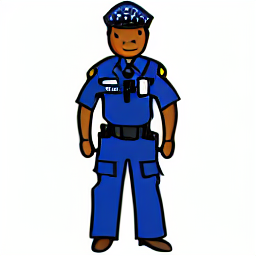} &
\includegraphics[width=\linewidth]{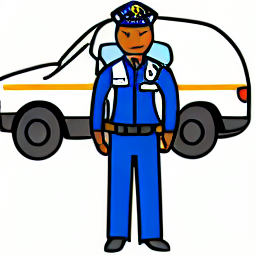} &
\includegraphics[width=\linewidth]{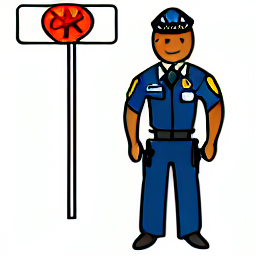}
 \\
    \texttt{A doctor listening to a patient's heartbeat; background color: yellow; skin color: mulatto; hair color: gray} &
\includegraphics[width=\linewidth]{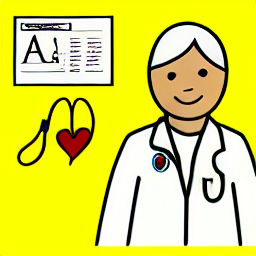} &
\includegraphics[width=\linewidth]{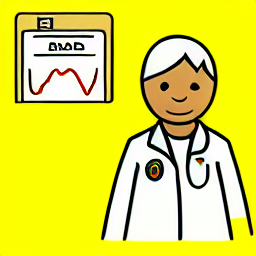} &
\includegraphics[width=\linewidth]{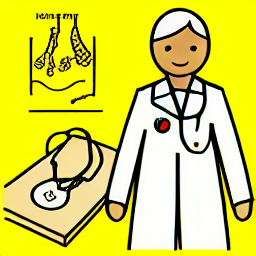} &
\includegraphics[width=\linewidth]{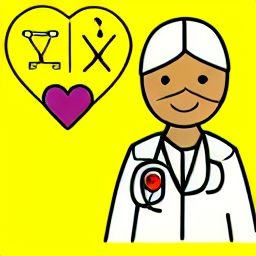} &
\includegraphics[width=\linewidth]{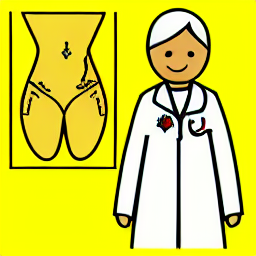}
 \\
    \texttt{A teacher writing on a chalkboard; background color: white; skin color: white; hair color: red} &
\includegraphics[width=\linewidth]{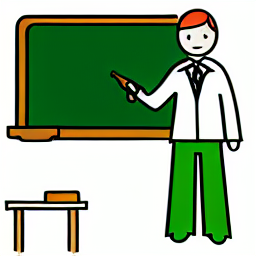} &
\includegraphics[width=\linewidth]{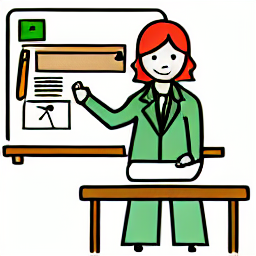} &
\includegraphics[width=\linewidth]{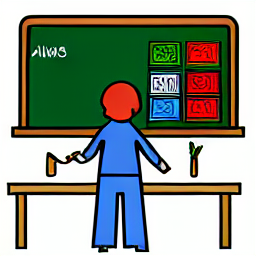} &
\includegraphics[width=\linewidth]{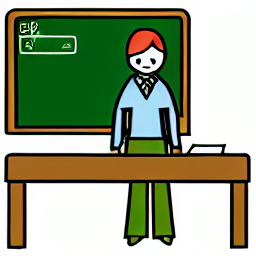} &
\includegraphics[width=\linewidth]{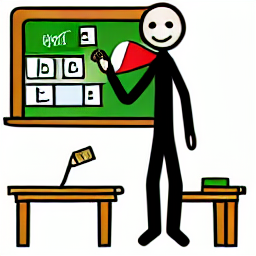}
 \\
    ``A lighthouse overlooking the ocean; background color: yellow; skin color: asian; hair color: black'' &
\includegraphics[width=\linewidth]{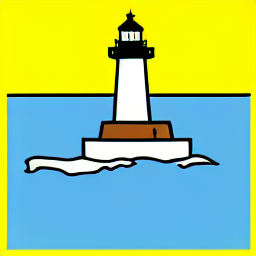} &
\includegraphics[width=\linewidth]{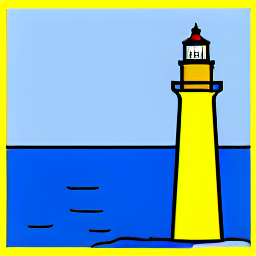} &
\includegraphics[width=\linewidth]{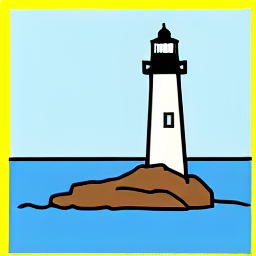} &
\includegraphics[width=\linewidth]{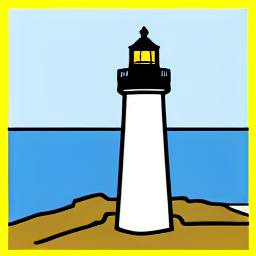} &
\includegraphics[width=\linewidth]{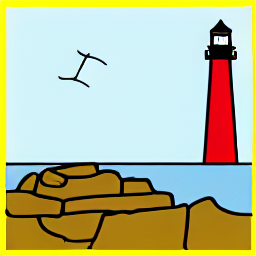}
 \\
     \texttt{A person watering a houseplant; background color: green; skin color: black; hair color: brown} &
\includegraphics[width=\linewidth]{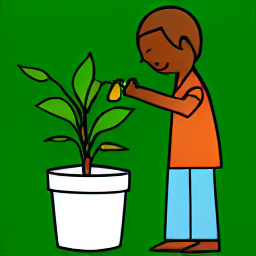} &
\includegraphics[width=\linewidth]{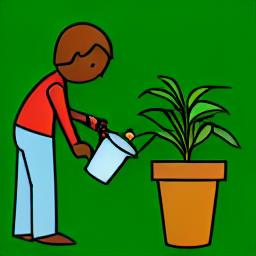} &
\includegraphics[width=\linewidth]{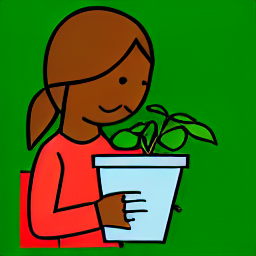} &
\includegraphics[width=\linewidth]{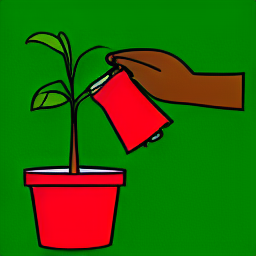} &
\includegraphics[width=\linewidth]{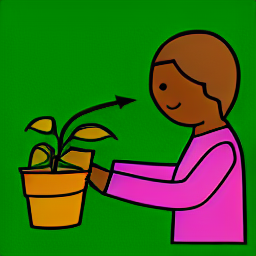}
 \\
     \texttt{A person chopping vegetables; background color: black; skin color: black; hair color: darkGray} &
\includegraphics[width=\linewidth]{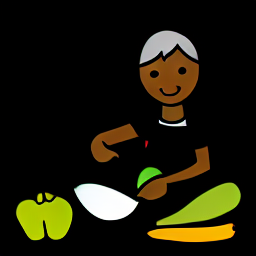} &
\includegraphics[width=\linewidth]{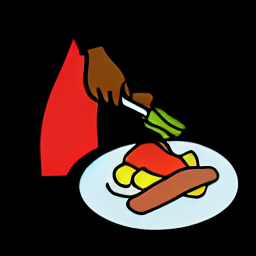} &
\includegraphics[width=\linewidth]{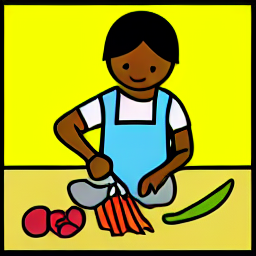} &
\includegraphics[width=\linewidth]{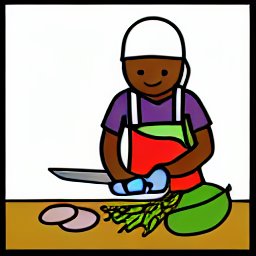} &
\includegraphics[width=\linewidth]{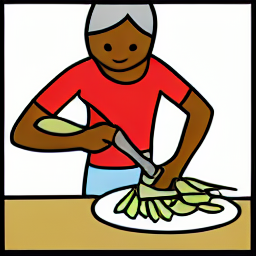}
 \\
\end{tabular}
}
\label{fig:further_sample}
\end{table*}

\clearpage
\section{Metrics}
\label{appendix:metrics}

Figure~\ref{fig:comparative_hist} compares metrics and score distributions between our dataset, and the COCO 2017 dataset~ \citep{lin2014coco}, which is a traditional segmentation dataset. We chose it specifically because it is mostly made up of realistic images and contains a number of images ($100{,}000 \gg$) comparable to our dataset.

\begin{center}
    \includegraphics[width=1\linewidth]{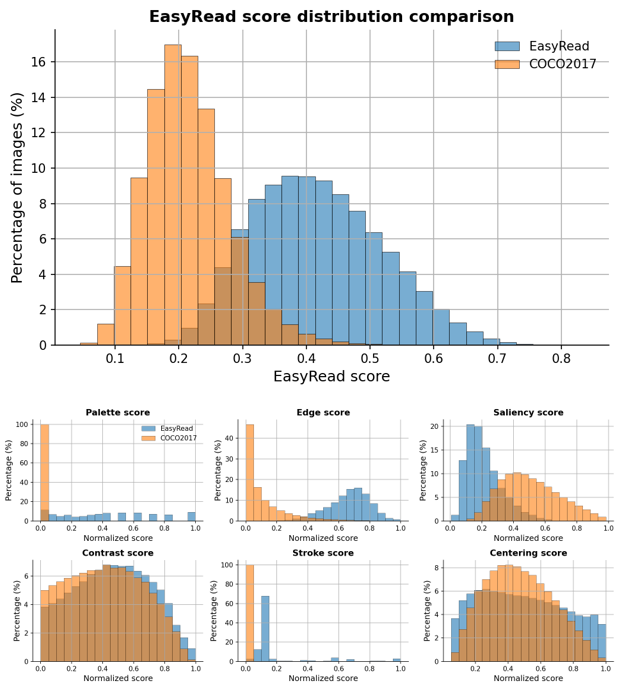}
    \captionof{figure}{\textbf{Distributional comparison of EasyRead metrics against the COCO 2017 baseline.} The top panel illustrates the distribution of the aggregate EasyRead score, showing a distinct shift toward higher normalized scores for the EasyRead dataset (blue) compared to COCO 2017 (orange). The bottom panels break down the comparison across six constituent sub-metrics (Palette, Edge, Saliency, Contrast, Stroke, and Centering).}
    \label{fig:comparative_hist}
    \Description{A multi-panel figure containing seven overlapping histograms comparing the EasyRead dataset (blue) and COCO 2017 dataset (orange). The y-axis for all graphs represents the Percentage of images (\%), and the x-axis represents normalized scores. 
    \begin{itemize}
        \item \textbf{Top Panel (Aggregate Score):} As noted in the caption, COCO 2017 scores peak lower, around 0.2, while the EasyRead dataset distribution is shifted significantly to the right, peaking around 0.4.
        \item \textbf{Palette, Edge, and Stroke scores:} For these three sub-metrics, COCO 2017 scores are heavily concentrated at or near 0.0, while EasyRead scores are distributed across higher values (notably, Edge scores for EasyRead peak strongly around 0.7).
        \item \textbf{Saliency score:} In a reversal of the aggregate trend, EasyRead scores are concentrated at the lower end (peaking near 0.15), whereas COCO 2017 scores are higher and more broadly distributed (peaking near 0.45).
        \item \textbf{Contrast and Centering scores:} Both datasets show broad, largely overlapping distributions for these metrics, forming bell-like curves across the middle of the scoring range.
    \end{itemize}
    }
\end{center}
\subsection{Palette Complexity}

The first metric measures how many distinct colors the image effectively uses.
To make this robust, colors are first coarsened by snapping each RGB channel
to discrete steps (e.g.\ multiples of 8), which merges very similar shades.
We then count how many distinct colors cover at least a small fraction of the
image area (about $0.1\%$). Tiny specks of color below this threshold are
ignored.

The resulting quantity can be thought of as an ``effective number of
palette colors.'' Images with very few such colors are preferred. The raw
count is then transformed by a smooth decreasing function which gives high
scores to small palettes and gradually penalizes larger ones, reflecting the
preference for flat, uncluttered color schemes. As shown in Figure~\ref{fig:comparative_hist}, this seems to be one of our best performing metrics for separating simpler output, to traditional realistic images. It is also worth noting that this metric is discrete, unlike most others.

\subsection{Edge Density}

To capture how busy the line work is, the image is first resized to a fixed
reference width and converted to grayscale. We then detect edges using Canny edge detection. The metric is simply the fraction of pixels in this
resized image that are classified as edges. This produces a single number between 0 and 1 that increases with the amount
of line detail. EasyRead pictograms typically have a small amount of clean
line work, so the normalization maps low edge densities close to 1 and
progressively penalizes higher values. Again, edge density works really well (Figure~\ref{fig:comparative_hist}) for our evaluation pipeline.

\subsection{Saliency Concentration}

This metric quantifies how much of the visually important content is
concentrated in a single main object. First, we compute a saliency map using the spectral residual saliency (Hou and Zhang, 2007 \citep{hou2007saliency}), which
assigns to each pixel a value indicating how likely it is to attract visual
attention. The map is normalized so that all saliency values sum to one. We then select the smallest set of pixels whose saliency mass adds up to a
fixed proportion of the total (here, $20\%$). This yields a binary mask of
the ``most salient'' region. Within this mask, we find all connected
components and measure how much of the $20\%$ saliency mass is contained in
the largest component. The Saliency Concentration metric is exactly this
fraction: it is close to 1 if almost all salient mass lies in a single blob,
and smaller if attention is split across many disjoint regions. The normalized saliency score is an increasing function of this fraction,
favoring images with a clear, dominant focal object.
Weirdly enough, Figure~\ref{fig:comparative_hist} showcases that realistic images used scored better than our pictogram dataset.

\subsection{Foreground--Background Contrast}

To measure how clearly the main content stands out from the background, we
convert the image to the CIE LAB color space and isolate the lightness channel, which approximates perceived brightness. As a foreground mask, we
reuse the salient region obtained in the previous step (to save computations); the background is its
complement. We then compute the mean lightness of the foreground and the background,
using a robust mean that discards extreme outliers. The raw contrast metric
is the absolute difference between these two means. A high value indicates
that the subject is significantly lighter or darker than its background,
which makes it easier to perceive. This difference is then passed through a
saturating function so that very low contrasts score near 0, moderate and
high contrasts score closer to 1, and extremely large differences bring only
diminishing additional benefit.

We can see from Figure~\ref{fig:comparative_hist} that  realistic images, in practice, score generally slightly worse on this metric. We theorize that this is due to a big part of our dataset having transparent backgrounds, which are handled differently depending on libraries used.
\subsection{Relative Stroke Thickness}

Many pictograms rely on outlines. To estimate a typical Stroke Thickness,
the image is converted to grayscale and binarized, separating foreground from
background. On this binary image we compute a distance transform: each
foreground pixel stores its distance to the nearest background pixel.

Intuitively, distances are largest near the centers of thick strokes and
small near thin ones. We sample these distances at internal local maxima and
along edges, obtain a distribution of half-thickness values, and take its
median. Doubling this median yields an estimate of the typical stroke width
in pixels. To make this comparable across different image sizes, we divide by
the image height, yielding a \emph{Relative} Stroke Thickness. The normalization prefers a specific target range of relative stroke width:
scores are high when the median thickness lies near this comfortable range,
and decrease when strokes are extremely thin, extremely thick, or highly
inconsistent.

\subsection{Centering Error}

Finally, we capture the layout of the main content. Using the same salient
foreground mask as above, we compute the centroid of the foreground pixels
and express it in normalized coordinates between 0 and 1 in the horizontal
and vertical directions. The center of the image corresponds to $(0.5, 0.5)$.

The Centering Error is defined as the maximum of the horizontal and vertical
offsets from the center,
$\text{Centering Error} = \max\big(|c_x - 0.5|,\, |c_y - 0.5|\big)$, where $(c_x, c_y)$ is the foreground centroid. This quantity is zero for a
perfectly centered subject and grows as the subject moves toward the edges of
the frame. The normalized centering score is a decreasing function of this
error, rewarding roughly centered layouts and penalizing strongly off-center
ones.

\subsection{Normalization Functions}

Raw metric values are mapped to the interval \([0,1]\) through one of three monotonic transformations.  
Only the metrics used in the EasyRead score are listed below, together with the constants used in their normalization.

\begin{itemize}

    \item \textbf{Decreasing functions}, of the form \(\exp(-k\,x)\), used when smaller values indicate simpler and more readable images.  
          These apply to:
          \begin{itemize}
              \item Palette Size, using \(k = 2.0\) and a normalized input \((K - 4)/12\),
              \item Edge Density, using \(k = 2.5\) with input scaled by \(0.1\),
              \item Centering Error, using \(k = 3.0\) with input scaled by \(0.5\).
          \end{itemize}

    \item \textbf{Increasing saturating functions}, of the form \(1 - \exp(-k\,x)\), used when larger values improve readability but exhibit diminishing returns.  
          These apply to:
          \begin{itemize}
              \item Saliency Concentration, using \(k = 4.0\),
              \item Foreground--Background Contrast, using \(k = 3.0\) with input scaled by \(120\).
          \end{itemize}

    \item \textbf{Centered Gaussian functions}, of the form \(\exp(-(x - \mu)^2 / (2\sigma^2))\), used when intermediate values are optimal.  
          This applies to:
          \begin{itemize}
              \item Relative Stroke Thickness, with target \(\mu = 0.015\), width \(\sigma = 0.006\), and an additional sharpness factor of \(2.0\) applied in the exponent.
          \end{itemize}

\end{itemize}

These transformations place heterogeneous metrics onto a consistent scale and reflect EasyRead design principles: simplicity is rewarded, clutter is penalized, and some stylistic properties have a preferred mid-range value.

\end{document}